\documentclass[natbib]{svjour3}                     
\smartqed  
\usepackage{graphicx}
\usepackage{epstopdf}
\usepackage{natbib}
\begin{document}

\title{Collisional and Radiative Processes in Optically Thin Plasmas
}



\author{Stephen Bradshaw         \and
        John Raymond 
}


\institute{S. J. Bradshaw \at
              Department of Physics and Astronomy, Rice University, 6100 Main St., Houston, TX 77005 \\
              \email{stephen.bradshaw@rice.edu}           
           \and
           J. Raymond \at
              Center for Astrophysics, 60 Garden St., Cambridge, MA  02138
}

\date{Received: date / Accepted: date}

\maketitle

\begin{abstract}
Most of our knowledge of the physical processes in distant plasmas is obtained through measurement of the radiation they produce.  Here we provide an overview of the main collisional and radiative processes and examples of diagnostics relevant to the microphysical processes in the plasma.  Many analyses assume a time-steady plasma with ion populations in equilibrium with the local temperature and Maxwellian distributions of particle velocities, but these assumptions are easily violated in many cases.  We consider these departures from equilibrium and possible diagnostics in detail.
\keywords{Microphysical processes}
\end{abstract}

\section{Introduction}
\label{intro}
Radiation is often the dominant cooling mechanism for optically thin astrophysical plasmas, which means that it determines the energy budget. It also provides most of the diagnostics for plasma parameters such as density, temperature and composition. It is therefore necessary to understand the dominant collisional and radiative processes in the plasma in order to answer astrophysical questions about the heating or energy dissipation in the plasma. In most cases, the radiation arises from collisions between electrons and ions, but interactions of electrons with a magnetic field or radiation field can also be important. 

The subsections of the introduction briefly summarize the processes that dominate
in most astrophysical settings, including the wavelength ranges
where they are observed and their identifying signatures.  In this section we
emphasize radiative signatures relevant to microphysical plasma processes,
such as differences between electron and ion temperatures, turbulence,
and non-Maxwellian velocity distributions.

The most detailed diagnostics for the physical parameters of plasmas and the microphysical processes taking place are generally based upon atomic and molecular lines and continua. In Section~\ref{atmmoldiag} we turn to a discussion of radiative processes and the diagnostics that are available. In Section~\ref{otemlines} we present the theory of  line formation in the coronal approximation and describe the dominant collisional and radiative processes. Section~\ref{ionization} comprises a discussion of the factors that influence the charge state, including the key ionization and recombination processes, the charge state in temperature equilibrium and the circumstances under which the charge state can become decoupled from the local temperature. The microphysics that arise when the electron distributions exhibit strong departures from Maxwellian are introduced in Secton~\ref{NMED}; we review the kinetic equations that describe the evolution of the distribution function and the different formalisms that have been adopted for handling collisions. In addition we address the consequences for the heat flux in terms of saturation and de-localization, and for the excitation and ionization rate coefficients which affect the ionization state and, in turn, the radiative losses. The optically-thin radiative loss function itself is the subject of Section~\ref{RLF} together with its dependence on the ionization state and the electron distribution. In Section~\ref{signatures} we return to a detailed review of the observational signatures and diagnostics that provide evidence for the importance of non-equilibrium ionization and non-Maxwellian electron distributions in the solar atmosphere. Finally, we present a summary of our review and look to the future in Section~\ref{summary}.

\subsection{Bremsstrahlung}
\label{brems}

Bremsstrahlung is continuum radiation produced by an electron
when it is accelerated in the electric field of an ion.  The
spectral shape is $P_{\nu} \sim exp(-h\nu /kT)$.  The mechanism
is well understood from basic electromagnetic theory \citep{rybicki}, but
relativistic corrections are needed for very high temperatures
and photon energies \citep{nozawa}.
Bremsstrahlung dominates the X-ray continua of many astrophysical
sources, though the continua due to radiative recombination
and 2-photon processes should not be ignored \citep{rs77},
and there could be a contribution from
synchrotron emission in young SNRs.

Bremsstrahlung emission in the X-rays generally arises from thermal plasmas,
but bremsstrahlung is also seen from beams of non-thermal electrons
in solar flares \citep{kontar11}.  Bremsstrahlung emission is also
referred to as free-free emission, particularly when observed at
longer wavelengths.  For example, free-free emission is observed
from planetary nebulae and H II regions in the radio,
and it is especially valuable as a measure of the ionizing flux
from the central star, because it is unaffected by reddening.

The signature of bremsstrahlung emission is a smooth continuum with
an exponential cutoff at $\rm h \nu \sim kT$.  For normal astrophysical
abundances it will be accompanied (and energetically dominated) by
spectral line emission unless the temperature is so high that the
abundant elements are ionized to their bare nuclei.

\subsection{Synchrotron and Cyclotron Emission}

The emission from electrons gyrating in a magnetic field
can be accurately predicted from electromagnetic theory \citep{rybicki}.
Relativistic electrons dominate the radio and X-ray synchrotron
emission from SNRs, the Galactic Halo, AGN and jets
from X-ray binaries.  Non-relativistic cyclotron emission
can be important in the solar corona and in accreting
magnetic white dwarfs.

Synchrotron emission dominates the radio emission of
supernova remnants, and in the fast shocks in young SNRs
it produces narrow filaments of X-ray emission.  The sharpness of the X-ray filaments
is used to derive limits on the diffusion coefficient for energetic particles
in the acceleration region \citep{long} and show that
the magnetic field is amplified well beyond the values
expected for compression in the shock \citep{vinklaming,
bamba}.  

Cyclotron and synchrotron emission are highly polarized,
but turbulence randomizes the field directions and Faraday
rotation can change the polarization direction and depolarize
the emission from an extended region. \cite{bykov} demonstrate
how turbulence will affect the X-ray polarization on small
scales, and \cite{dickel} have shown that the radio polarization
indicates radial, rather than tangential magnetic fields near the edge
of Tycho's SNR.  Polarization maps in the radio provide a
unique method for observing the turbulent structure of the
galactic magnetic field \citep{haverkornheesen}.

Synchrotron emission dominates the radio and X-ray
spectra of pulsar wind nebulae (PWNe), jets from AGN
and gamma-ray bursts.  It is straightforward to determine
the power law slope of the emitting electrons from the
slope of the spectrum.  The ambiguity between magnetic
field strength and the number of emitting electrons can
sometimes be resolved based on spectral breaks due to
optical depth or synchrotron cooling.

The emission and absorption occur between quantized Landau levels in
the solar corona at radio wavelengths \citep{dulk}, in magnetic
cataclysmic variables in the optical,
and in accreting neutron stars in the X-ray.  The emission
at harmonics of the cyclotron frequency can be used to determine
the magnetic field strength. 
The lowest harmonics often are optically thick and the
higher ones optically thin.  At the transition, the radiation
can be strongly polarized.  For example,
\cite{brosiuswhite} used radio measurements above the solar limb
to obtain the magnetic field strength above a sunspot.

The signatures of synchrotron emission from relativistic electrons
are a power law spectrum and a substantial polarization fraction.
Gyro emission from non-relativistic thermal electrons typically
shows a spectral peak corresponding to a modest harmonic of the
cyclotron frequency, with substantial polarization.

\subsection{Compton and Inverse Compton Heating and Cooling}
\label{compton}

The interaction between a photon and an electron can transfer
energy either way.  As for bremsstrahlung and synchrotron emission,
the physical process is well understood \citep{rybicki}.  Hot
plasma above an accretion disk
will experience Compton heating by hard X-rays from the central
source and Compton cooling by softer photons from the disk.
Energetic electrons can interact with synchrotron photons
produced by the same electron population (synchrotron self compton
emission).

In supernova remants, the energetic electrons can produce
TeV gamma rays by inverse Compton interaction with the
cosmic microwave background (CMB) or with locally enhanced IR
or optical radiation.  It is currently debated whether
the gamma ray emission observed from several SNRs arises
from inverse Compton emission by energetic electrons or
from decay of pions produced by interactions between
cosmic ray protons and dense ambient plasma. Consideration of
the lower energy gamma rays observed by FERMI can help to resolve
the ambiguity.
Inverse Compton gamma ray observations provide
at least a lower limit to the maximum energy of the accelerated
electrons, and they provide the number of energetic electrons.
The latter, in combination with the synchrotron X-ray emission
also determines the magnetic field strength.

The signature of inverse compton emission is a high energy continuum.
It is most important when the radiation field is strong and the plasma density
and magnetic field are relatively low.

\subsection{Plasma Emission}
\label{plasma}

Plasma emission is one of the most direct links between microphysical
processes in a plasma and Earth-based observations.  It is most
commonly seen in the form of type II and type III radio bursts
from the Sun.  In either case, the
radio emission is produced by a several step process.  A beam of electrons
produced in a flare or at a shock front penetrates into thermal
plasma, giving an unstable bump-on-tail velocity distribution.  That
distribution produces Langmuir waves at the plasma frequency
as it flattens into a stable distribution.  The Langmuir
waves can produce backscattered Langmuir
waves and ion acoustic waves, and subsequent interactions between the
beam-driven waves and these secondary waves produce radio emission
at the plasma frequency of 9$n_e^{1/2}$ kHz and twice the plasma frequency
\citep{pick}. \cite{schmidtcairns} present an analytical
formalism for the type II radiation from a shock.  The emission
tends to be strongest at nearly perpendicular shocks \citep{cairns11}.

Though plasma emission has been reported from the coronae of a few
active stars, it is seldom observed from astrophysical sources.
Most objects bright enough to observe are very dense and optically
thick.  Type II emission is almost certainly produced by shock waves in
supernova remants, but radiation at the kHz frequencies  given
by the density of the ISM does not reach Earth.

Since the emission is at the plasma frequency or first harmonic,
the measured frequency directly gives the density in the emitting
region.  The drift rate of the frequency gives the shock speed
for a type II burst if the density structure is known.  However,
shock speeds inferred from type II drift rates do not agree well
with shock speeds measured by coronagraphs \citep{mancuso},
either because an inappropriate density structure is assumed or
because different parts of the CME shock emit as the CME evolves
due, for instance, to selection of a particular angle between
the field and the shock where emission is efficient.

The signatures of plasma emission from the solar corona are enormous
brightness temperatures and narrow bands of emission near the plasma
fequency.

\subsection{Dust Emission}

In many cases a plasma is optically thin to radiation from dust,
even though the emission from individual dust grains is optically
thick at some wavelengths.  Grains in relatively hot plasmas
that are heated to temperatures of order 10-100 K emit at
sub-millimeter and infrared wavelengths.  The spectrum is
a blackbody modified by the opacity of the grain, so it may
contain features such as the silicate bump at 9.7 microns that can
reveal the nature of the grain material.

Behind the fast shock wave of a supernova remnant, dust
is heated to temperatures around 100 K, mainly by collisions
with electrons, even as it is gradually eroded by sputtering
due to collisions with ions.  Infrared emission 
by dust can be the
main radiative energy loss from shock waves faster than about
300 km/s \citep{arendt}.  The spectrum and the intensity falloff
behind the shock can be used to infer the post-shock density
and the destruction rate of the dust \citep{williams06, williams08,
williams11, sankrit}.

Dust also absorbs and scatters light at optical, UV and X-ray
wavelengths.  The wavelength dependence of the absorption,
in particular the 2200 \AA\/ feature, in combination with
the IR emission spectrum, is used to infer the
size distribution and composition of the dust \citep{draine}.
If the dust column density is fairly high, a detectable
halo of X-rays appears around a bright X-ray point source
\citep{smithedgar}, from which one can derive the grain size distribution
and the location of the grains along the line of sight.

The signature of dust emission is a blackbody-like spectrum
at IR or sub-millimeter wavelengths, sometimes with discrete
features due to silicates, polycyclic aromatic hydrocarbons
(PAHS) or other features.  The brightness is proportional to
the amount of dust and grain size distribution.  It is also sensitive 
to temperature, which in turn is sensitive to either the radiation 
field that heats the grains or the density and temperature of the i
gas in which they are immersed.

\subsection{Ion-Ion collisions}
\label{ionion}

While most of the radiation detectable at Earth is
produced by electrons, energetic collisions between
ions produce observable gamma rays. These are most
clearly seen during solar flares, when energetic
ions strike the dense gas of the chromosphere to
produce broad and narrow nuclear de-excitation lines,
positrons that subsequently annihilate to produce
0.511 MeV photons, and neutron capture lines
\citep{vilmer11}.  Gamma ray spectra from RHESSI
and other instruments can be used to infer the
composition of both the chromosphere and the
energetic ions, the spectral shape of the accelerated
particles and their energy content at MeV energies.

At higher energies, cosmic rays can collide with
nuclei in the ambient gas to produce pions, which
can decay into gamma rays.  Though it is often
difficult to tell whether TeV gamma rays are produced
by pion decay or inverse Compton interaction between
ambient photons and energetic electrons, observations
of supernova remnants with the ground-based arrays
H.E.S.S, MAGIC and VERITAS, and with the FERMI
satellite, offer constraints on
the acceleration of hadrons in stong shock waves.
The nature of the gamma ray emission from many SNRs
is still under debate, but
the gamma rays from some old SNRs interacting with
dense clouds can be attributed to pion decay
(e.g., \cite{uchiyama} FERMI observations of W44).

\section{Atomic and Molecular Spectral Line Diagnostics}
\label{atmmoldiag}

Atomic spectral lines can appear in emission or absorption. Emission lines usually arise following excitation by electron impact or recombination into an excited level, though they can also be produced by ion impact \citep{laming} or photoabsorption \citep{noci}. Cooling by emission of atomic or molecular lines often dominates the energy budget of the plasma, and the intensities of the spectral lines provide powerful diagnostics for the physical parameters of the plasma. This Section provides an overview of atomic emission line diagnostics, and Section~\ref{otemlines} provides a rigorous discussion of the line formation process.

\begin{figure}
\centering
\includegraphics[width=1.0\textwidth]{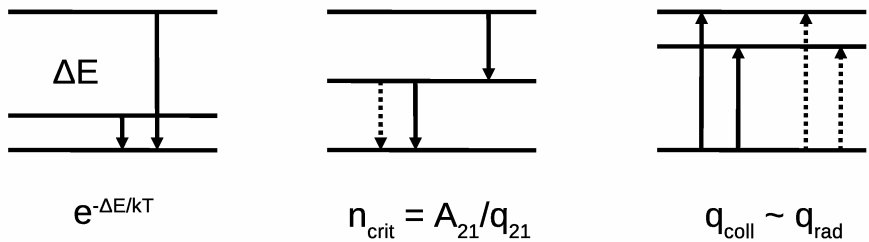}
\caption{Temperature diagnostics are generally based on the ratio of Boltzmann factors, $\exp^{-\Delta E/k_BT}$, in the excitation rates of two spectral lines (left diagram). Density diagnostics can be based on the competition between radiative decay and collisional de-excitation when the Einstein $A$ value is comparable to the density times the collisional rate coefficient (middle panel) or on the relative contributions of collisional excitation and radiative excitation (right panel).}
\label{schematic}
\end{figure}

Intensity ratios of lines within a single ion can be used to infer the electron temperature and density of the gas. Electron temperature diagnostics generally hinge on the Boltzmann factor,
$\exp^{-\Delta E/k_BT}$, where $\Delta E$ is the energy difference between the two upper levels (Figure~\ref{schematic} left diagram). Such a ratio works best for $\Delta E \sim k_BT$, so that optical line ratios are effective for $T$ around $10^4$ K, where $\Delta E \sim k_BT \sim 1$~eV. UV line ratios are effective around $10^5$ K and X-ray line ratios above $10^6$ K. Often the desirable spectral lines lie at much different wavelengths, so that it is hard to obtain a ratio with a single instrument, but the technique has been applied to solar spectra \citep{david}.

The density can be inferred from ratios involving a metastable level. The population of that level will be small at low densities.  It approaches a constant value given by the statistical weight and Boltzmann factor above a critical density $n_{crit} =A_{21}/q_{21}$, where $A_{21}$ is the Einstein $A$ value and $q_{21}$ is the de-excitation rate coefficient. The ratio of a line which involves the metastable level to a line which does not will be sensitive to density (Figure~\ref{schematic} center diagram). Because the $A$ values increase rapidly with transition energy and $q$ values decline, $n_{\mbox{crit}}$ increases rapidly from values around $10^2$ to $10^4$ $\rm cm^{-3}$ for optical forbidden lines to $10^8$ to $10^{10}$ $\rm cm^{-3}$ for UV lines and $10^{11}$ to $10^{15}$ $\rm cm^{-3}$ for X-ray lines.
 
A less commonly used density diagnostic takes advantage of the fact that some lines formed in the solar corona include both collisionally excited and radiatively excited components. The ratio of those components is proportional to the density and perhaps plasma velocity and line width \citep{noci}. It is interesting to note that for an ion X, the ratios indicated in the center and right panels of Figure~\ref{schematic} give $<n_X n_e^2/(n_{crit}+n_e)>/<n_X n_e/(n_{crit}+n_e)>$ and $<n_X n_e>/<n_X W>$, respectively.  Here $W$ is the dilution factor of the radiation (Section~\ref{RadProc}). Thus different density estimates are differently weighted averages that do not necessarily agree. In principle, comparison of differently weighted averages could yield unique information about the distributions of electron density and density of the diagnostic ion within the observed volume, but that requires very good accuracy for both diagnostics \citep{leeetal08}.  

Ratios of emission or absorption lines of different elements can also be used to derive the relative elemental abundances. In practice that is often tricky because in many cases only 1 or 2 ions of each element can be observed, so the ionization state of each element must be accurately known. This usually requires a model that involves ionization and recombination rates, each
having perhaps a 20\% uncertainty, and it often involves an assumption of ionization equilibrium (Section~\ref{IonEq}) that may not be justified (Sections~\ref{neq} and \ref{IonNEq}).  These difficulties are somewhat mitigated if one can use ions such as He-like and H-like ions that dominate the ionization distribution over broad temperature ranges.

The profiles of optically thin emission or absorption lines provide a direct measurement of the velocity distribution of atoms, molecules or ions along the line of sight. Therefore, they provide good diagnostics for the ion kinetic temperatures, turbulence and in principle non-Maxwellian velocity distributions (Section~\ref{NMED}), though there may be ambiguities among the different interpretations.

Line profiles directly give ion temperatures when bulk motions do not dominate. In low density regions of the solar corona, the line widths of oxygen ions exceed those of hydrogen, indicating
that the kinetic temperature of O is more than 16 times that of H \citep{kohl, cranmer, frazin}.

Collisionless shock waves are another good example of the application of line profile diagnostics. Neutral hydrogen that passes through a strong shock does not feel the collisionless shock itself, but finds itself immersed in the hot post-shock flow. Diagnostics based on Balmer line profiles from these shocks are discussed in \cite{bykov12}. Most observed line profiles can be fit with a Gaussian or a sum of Gaussians, so they are consistent with Maxwellian distributions. The broad H$\alpha$ profile of a bright knot produced by a 2000 km/s shock in Tycho's supernova remnant is not Maxwellian, suggesting either a power-law tail or a pickup-ion contribution, though an interpretation as a sum of Maxwellian contributions cannot be excluded \citep{raymond10}. 

Line profiles can be directly used to determine the level of turbulent velocity fluctuations if thermal and bulk velocities do not dominate. Comparison of lines from elements of different masses can help to resolve the ambiguity between thermal and turbulent line widths. Line widths have been used to estimate the level of turbulence in reconnection current sheets during solar
eruptions \citep{bemporad}. Another application has been study of turbulence in interstellar gas using the Velocity Coordinate Spectrum method to combine line profiles and their spatial variations \citep{lazarian}. These statistical methods, along with methods based on polarization \citep{burkhart} can reveal the turbulence spectrum and whether the turbulence is subsonic. 

\section{Optically-Thin Emission Lines}
\label{otemlines}

\subsection{Line Formation}

High temperature $\left(>10^6\mbox{~K}\right)$ and low density $\left(<10^{13}\mbox{~K}\right)$ astrophysical plasmas are optically-thin to visible, EUV and X-ray radiation. Photons at these wavelengths are generally able to propagate through these environments unhindered by opacity effects, such as absorption and re-emission, and scattering, and therefore retain a record of the plasma conditions at the site of emission. Most of the radiation in the region of $10^6$~K is due to the emission of photons by electron transitions in ions, giving rise to spectral lines. The radiated power per unit volume, commonly referred to as the {\it emissivity}, depends on: (a) the number of ions that are present; and (b) the fraction of those ions in the excited state that corresponds to the transition. For a given transition \citep[in the notation of][]{Mason1994}:

\begin{equation}
P\left(\lambda_{j,i}\right) = N_j\left(X^{+m}\right) A_{j,i} \Delta E_{j,i} \mbox{~~~~[erg~cm$^{-3}$~s$^{-1}$]},
\label{eqn3sb}
\end{equation}

\noindent where $N_j\left(X^{+m}\right)$~[cm$^{-3}$] is the number density of ions of charge $+m$ in excited state $j$, $A_{j,i}$~[s$^{-1}$] is the Einstein coefficient and $\Delta E_{j,i}$ is the energy of the emitted photon. The quantity $N_j\left(X^{+m}\right)$ can be rewritten as a series of ratios that can be measured observationally or experimentally, or calculated theoretically \citep{Mason1994}. The total energy flux due to the transition, at a distance $R$ from the emitting volume of plasma, can be found by integrating the emissivity over the volume and dividing by the surface area of the sphere with radius $R$:

\begin{equation}
I\left(\lambda_{j,i}\right) = \frac{1}{4 \pi R^2} \int_V P\left(\lambda_{j,i}\right) dV \mbox{~~~~[erg~cm$^{-2}$~s$^{-1}$~sr$^{-1}$]}.
\label{eqn5sb}
\end{equation}

\subsection{The Coronal Model}

A convenient approximation for optically-thin plasmas, such as the solar corona, allows a decoupling of the processes that determine the excitation state from those that determine the charge state. This can be justified by noting that changes in the energy level populations of the emitting ions occur far more frequently than changes in the charge state. The processes that determine the excitation state are discussed in this Section and those that determine the charge state are discussed in Section~\ref{ionization}. In optically-thin plasmas energy levels become populated by electron collisional excitation from the ground-state ($g$) of each ion, and they become depopulated by spontaneous radiative decay. It is assumed that timescales of photon absorption and electron collisional de-excitation are far longer. This is called the {\it coronal model} approximation and in statistical equilibrium the number of collisional transitions from the ground-state $g$ to the excited state $j$ must be equal to the number of spontaneous radiative decays back to the ground-state.

\begin{equation}
N_g\left(X^{+m}\right) N_e C_{g,j}^e = N_j\left(X^{+m}\right) A_{j,g} \mbox{~~~~[cm$^{-3}$~s$^{-1}$]}.
\label{eqn6sb}
\end{equation}

\noindent $C_{g,j}^e$~[cm$^3$~s$^{-1}$] is the electron collisional excitation rate coefficient between the ground-state and level $j$. If collisions are relatively infrequent then $A_{j,g} >> N_e C_{g,j}^e$ and it follows that $N_g\left(X^{+m}\right) >> N_j\left(X^{+m}\right)$. There are many more ions in the ground-state than in excited states. In a typical transition at EUV wavelengths $A_{j,g}=10^{10}$~[s$^{-1}$] and $N_e C_{g,j}^e = 1$~[s$^{-1}$] and so for every collisional excitation there is an almost immediate radiative decay to satisfy the requirements of statistical equilibrium. We note that Equation~\ref{eqn6sb} pertains to 2-level atoms, but radiative cascades from higher levels, following excitation or recombination, may dominate under particular circumstances, such as transitions from Fe~XVII~3s levels \citep{Beiersdorfer2004}. 

The statistical equilibrium relationship given in Equation~\ref{eqn6sb} and the fact that $\frac{N_g\left(X^{+m}\right)}{N\left(X^{+m}\right)} \approx 1$ leads to an expression for the emissivity in terms of the collisional excitation rate:

\begin{equation}
P\left(\lambda_{j,g}\right) = \frac{N\left(X^{+m}\right)}{N(X)} \frac{N(X)}{N(H)} \frac{N(H)}{N_e} C_{g,j}^e \Delta E_{j,g} N_e^2.
\label{eqn7sb}
\end{equation}

\noindent The spectral line intensity is proportional to $N_e^2$ as expected.

\subsection{Collisional Processes}
\label{ColProc}

The rate at which collisional transitions occur depends on the interaction cross-section presented to incident particles by the target and on the flux of incident particles. The flux of incident particles can be written:

\begin{equation}
F = nvf(E)dE \mbox{~~~~[particles~cm$^{-2}$~s$^{-1}$]},
\label{eqn8sb}
\end{equation}

\noindent where $n$ is the number density of particles, $v$ is the incident particle velocity, $E$ the kinetic energy of the incident particles and $f(E)$ the particle distribution function. Since particle-particle interactions are mostly via collisions then it is common to assume that the distribution function is a collisionally relaxed Maxwellian of the form:

\begin{equation}
f(E) = 2 \sqrt{\frac{E}{\pi}} \left(\frac{1}{k_B T}\right)^\frac{3}{2} \exp \left(-\frac{E}{k_B T} \right) \mbox{~~~~[particles~erg$^{-1}$]}.
\label{eqn9sb}
\end{equation}

\noindent The electron collisional excitation rate coefficient is found by integrating the electron distribution function over the interaction cross-section.

\begin{equation}
C_{i,j}^e = \int_{\Delta E}^\infty Q_{i,j} v f(E) dE \mbox{~~~~[cm$^3$~s$^{-1}$]}.
\label{eqn10sb}
\end{equation}

\noindent $\Delta E$ is the energy difference between level $i$ and $j$, and this is the lower limit to the integral because an incident particle must have at least this much energy in order to excite the transition. $Q_{i,j}$~[cm$^2$] is the interaction cross-section. In simplified form

\begin{equation}
C_{i,j}^e = \frac{8.63\times10^{-6}}{\omega_i \sqrt{T}} \Upsilon_{i,j} \exp \left(-\frac{\Delta E}{k_B T}\right),
\label{eqn13sb}
\end{equation}

where $\omega_i$ is the statistical weight of level $i$, which is the number of different spin and angular momentum states that have energy $E_i$ (the number of degenerate states in energy $E_i$), and $\Upsilon_{i,j}(T)$ is the thermally averaged collision strength \citep{Mason1994}. $\omega_i=2n_q^2$ for hydrogen (where $n_q$ is the principle quantum number).

\subsection{Radiative Processes}
\label{RadProc}

Spontaneous radiative decay of electrons from excited states is the dominant depopulation mechanism in optically-thin plasmas. The generalised radiative decay coefficient is:

\begin{equation}
R_{j,i} = A_{j,i} \left( 1 + \frac{W}{\exp \left(\frac{\Delta E}{k_B T}\right) - 1}\right) \mbox{~~~~[s$^{-1}$]}.
\label{eqn14sb}
\end{equation}

The first term of Equation~\ref{eqn14sb} takes account of spontaneous emission. The second term accounts for the stimulated component of the emission in the presence of a background continuum radiation field, described by a Planck function. In the case of radiative decay in the solar corona the background radiation field would have a temperature of 5800~K, characteristic of the photosphere. $W$ is a dilution factor that describes the decay of the radiation field with radial distance, where:

\begin{equation}
W = \frac{1}{2} \left[ 1 - \left( 1 - \frac{R_0^2}{r^2} \right) \right].
\label{eqn15sb}
\end{equation}

In the case of the Sun, $R_0$ would be the solar radius and $r$ the distance from the centre of the Sun to the height in the atmosphere at which $W$ must be calculated. As $r \rightarrow \infty$ (e.g. sufficiently far above the surface that $r>>R_0$) then $W \rightarrow 0$ and the stimulated component of the emission can be neglected so that $R_{j,i} = A_{j,i}$. The stimulated component of the solar radiative flux is also negligible at far UV and shorter wavelengths; however, photoexcitation of UV lines such as the Lyman series and O~VI is very important beyond about 1.3 solar radii.

\section{The Charge State of a Plasma}
\label{ionization}

\subsection{Ionization and Recombination}

The charge state of the ions in a plasma is governed by the rate at which electrons are freed from their bound states and the rate at which free electrons are captured into bound states. Bound-free transitions are called {\it ionization} and free-bound transitions are called {\it recombination}. Collisional excitation and radiative decay occur on timescales far shorter than ionization and recombination timescales, and so these processes can be de-coupled from the excitation and decay processes. Ionization~(recombination) can then be considered to take place from~(to) the ground-state of the ion, though it is worth noting that at transition region densities (e.g. $n \approx 10^{10}$~cm$^{-3}$) ionization and recombination from metastable levels can become important \citep{Vernazza1979}. In optically-thin plasma, such as solar and stellar coronae, the important ionization processes are: collisional ionization; and excitation-autoionization. The important recombination processes are: radiative recombination; and dielectronic recombination.

\textbf{Collisional ionization}: as in the case of collisional excitation, the dominant process of ionization is by electron collisions (photo-ionization is negligible at the energies of interest). Where collisional excitation is generally due to electrons in the bulk of the distribution (e.g. a Maxwellian), ionization arises from electrons in the high-energy tail. Since the number density of electrons in the tail is relatively low then collisional ionization is relatively infrequent compared with collisional excitation. The process of collisional ionization can be written \citep[again employing the standard notation of][]{Mason1994}:

\begin{equation}
X_i^{+m} + e^- \rightarrow X_{i'}^{+m+1} + 2e^-.
\label{eqn16sb}
\end{equation}

\noindent The ion in the state $i$ loses an electron and a new ion is created in the state $i'$. The incident electron must have sufficient energy to free the bound electron and retain enough to remain unbound.

\textbf{Radiative recombination}: similarly to radiative decay, an important recombination process is the capture of an energetic free electron into a lower energy, bound state, leading to the emission of a photon. The radiative recombination process can be written:

\begin{equation}
X_{i'}^{+m+1} + e^- \rightarrow X_i^{+m} + \Delta E.
\label{eqn17sb}
\end{equation}

\textbf{Dielectronic recombination}: the dominant recombination mechanism at high temperatures, as shown by \cite{Burgess1964}. The dielectronic recombination process can be written:

\begin{equation}
X_{i'}^{+m+1} + e^- \rightarrow \left(X_{i''}^{+m}\right)^{**} \rightarrow X_i^{+m} + \Delta E.
\label{eqn18sb}
\end{equation}

\noindent Equation~\ref{eqn18sb} shows that an ion with $m+1$ missing electrons may capture a free electron into a particular outer energy level while simultaneously exciting an inner electron to a higher energy level {\it instead} of emitting a photon. The $\left(~\right)^{**}$ notation indicates a doubly excited state. The excited inner electron may then decay to its original level (or another, if low-lying fine structure states are available), with the emission of a photon, leaving the ion in a singly excited state because the captured electron remains in an outer energy level. At this point the recombination is complete. Dielectronic recombination is the dominant recombination mechanism for most ions at high temperatures, especially those with $\Delta n_q = 0$ transitions from the ground state. Dielectronic recombination can also be somewhat density dependent, because the emission of a photon often leaves the recombined ion in a highly excited state that can be ionized before it decays to the ground state.

\textbf{Excitation-autoionization}: if the two excited electrons in the second stage of Equation~\ref{eqn18sb} together have more energy than is needed to remove a single electron from the ground state, then the ion is energetically able to autoionize. This means that it can decay to the ground state with the ejection of one of the excited electrons:

\begin{equation}
\left(X_{i''}^{+m}\right)^{**} \rightarrow X_i^{+m} + e^-.
\label{eqn19sb}
\end{equation}

\noindent Note that the process described by Equation~\ref{eqn19sb} is the inverse process to the first stage of dielectronic recombination in Equation~\ref{eqn18sb}. The doubly excited ion has two choices: (1) emit a photon; or (2) autoionize (if the total energy of the excited electrons exceeds the threshold for ionization).

\textbf{Charge transfer}: charge transfer between ionized species and neutral hydrogen is not usually important in the solar corona, but it can modify the ionization state in astrophysical plasmas, especially cool plasma photoionized by a hard radiation field. The cross-section for resonant charge transfer is very large and this sometimes makes up for a low neutral fraction.

\subsection{The Charge State in Equilibrium}
\label{IonEq}

Ionization and recombination rate coefficients depend strongly on temperature and, to a somewhat lesser extent, on density. At higher temperatures the free electrons have a greater average kinetic energy and so are able to collisionally release even the strongly bound, inner electrons of the target ions. At lower temperatures the free electrons are less energetic and can be captured even into the low ionization energy, outer bound states of the ions. It is useful to observe that ions are typically found at a temperature such that the ionization potential is $\approx 5k_BT$ in equilibrium. A full set of ionization and recombination rate coefficients \citep[e.g][]{Arnaud1985,Arnaud1992,Mazzotta1998,Bryans2009,Dere2007} for a given element allows the distribution among the charge states for the ions of that element to be calculated as a function of temperature. We note here that published rate coefficients tend to be calculated assuming that the free electrons have relaxed into a Maxwellian electron distribution. We consider the consequences of the breakdown of this assumption and the calculation and the consequences of departures from an underlying Maxwellian in Sections~\ref{NMED}.

One may ask what proportion of helium atoms are neutral, singly ionized and doubly ionized at a particular temperature. This is the charge or ionization state of the element. At at a temperature of $10^6$~K helium is fully ionized and so the population fractions are: He~I~(neutral)~=~0.0; He~II~(singly-ionized)~=~0.0; and He~III~(doubly-ionized)~=~1.0. At $10^5$~K (adopting the ionization rates of Dere~2009 and the recombination rates of Mazzotta~et~al. 1997) the ionization state of helium is: He~I~=~0.0; He~II~=~0.131; and He~III~=~0.869. 13\% of helium is singly ionized and 87\% of helium is fully ionized at $10^5$~K. The population fractions for all the ions of a particular element must sum to 1.0 in order to conserve the particle number.

The population fraction for each ion peaks at the temperature at which the ionization and recombination rates are equal. More ionizations would act to deplete the ion population in favour of a higher charge state, and more recombinations would deplete the population in favour of a lower charge state. The ionization states for helium given above are only reached when the ionization state is in {\it equilibrium} with the electron temperature of the plasma. Strictly speaking, as $t \rightarrow \infty$ at $T=10^5$~K then $\mbox{He~I}\rightarrow 0.0$, $\mbox{He~II}\rightarrow 0.131$  and $\mbox{He~III}\rightarrow 0.869$. The reason for this is that collisional processes are not instantaneous. It takes a certain period of time for ionization and recombination events to arrange the ions into the charge states that correspond to the current electron temperature. As long as the ionization and recombination timescales are much shorter than the timescale on which the temperature changes then the ionization state can be considered in equilibrium with the temperature, and therefore depends only on the temperature. The break-down of this condition will be discussed in Section~\ref{neq}.

One consequence of de-coupling ionization and recombination from the processes of excitation and radiative decay is that one may assume ionization~(recombination) occurs from~(to) the ground state of the ion, and so the rate of change of the population fraction of a particular ion $i$ of element $X$ can be written:

\begin{equation}
\frac{dX_i}{dt}=n\left(I_{i-1} X_{i-1} + R_i X_{i+1} - I_i X_i - R_{i-1} X_i\right).
\label{eqn20sb}
\end{equation}

\noindent In the notation of Equation~\ref{eqn20sb} element $X$ might be helium and then $X_{i=0}$ would be neutral helium (He~I), and so forth. $n$~[cm$^{-3}$] is the electron number density, and $I_i$ and $R_i$ are the temperature dependent {\it total} ionization and recombination rate coefficients, respectively, with units [cm$^3$~s$^{-1}$]. In equilibrium $\frac{d}{dt}=0$ so that:

\begin{equation}
I_{i-1} X_{i-1} + R_i X_{i+1} = I_i X_i + R_{i-1} X_i.
\label{eqn21sb}
\end{equation}

\noindent The LHS of Equation~\ref{eqn21sb} comprises the processes that lead to the creation of ion $X_i$ (ionization from lower charge states and recombination from higher charge states). The RHS comprises the processes that lead to the destruction of $X_i$ (ionization to higher charge states and recombination to lower charge states). In equilibrium the principle of detailed balance implies that the rate of ionization to $X_i$ is equal to the rate of recombination from $X_i$, and the rate of ionization from $X_i$ is equal to the rate of recombination to $X_i$. This can be expressed in the form of two de-coupled equations:

\begin{equation}
I_{i-1} X_{i-1} = R_{i-1} X_i;
\label{eqn22sb}
\end{equation}

\begin{equation}
R_i X_{i+1} = I_i X_i.
\label{eqn23sb}
\end{equation}

\noindent The ionization state can then be fully specified subject to the final constraint:

\begin{equation}
\Sigma_{i=0}^Z X_i = 1.0,
\label{eqn24sb}
\end{equation}

\noindent where $Z$ is the atomic number of the element $X$. Making use of Equations~\ref{eqn22sb} and \ref{eqn23sb} it can be seen that:

\begin{equation}
X_{i-1} = \frac{R_{i-1}}{I_{i-1}}X_i \mbox{~~~~and~~~~} X_{i+1} = \frac{I_i}{R_i}X_i.
\label{eqn25sb}
\end{equation}

Given a set of ionization and recombination rate coefficients the ionization state can be calculated by choosing a suitable value for $X_i$. The most abundant ion $i$ of element $X$ is the one for which $I_i(T) \approx R_i(T)$ at the temperature of interest. The population fraction of this ion can then be assigned some arbitrary quantity $X_i=X'_i$ usually chosen to avoid computational overflow errors since the population fractions can vary over many orders of magnitude (this is not so much of an issue in the case of double-precision arithmetic). It is then straightforward to calculate $[X'_{i-1},X'_{i-2},...,X'_0]$ and $[X'_{i+1},X'_{i+2},...,X'_Z]$ recursively from Equation~\ref{eqn25sb} and find the true population fractions by normalising the values of $X'_{i}$ to 1.0:

\begin{equation}
X_i = \frac{X'_i}{\Sigma_{i=0}^Z X'_i}.
\label{eqn26sb}
\end{equation}

\subsection{Non-Equilibrium Charge States}
\label{neq}

In circumstances where the electron temperature has been held fixed for a long time {\it or} the temperature is changing slowly, then the ionization state of the plasma is in equilibrium and depends on the temperature only. A slowly changing temperature in the present context means that it changes more slowly than the timescales on which the processes that change the ionization state of the plasma operate. If the temperature change is sufficiently slow then collisions have ample time to arrange the charge states of the element such that they are in equilibrium with the temperature.

Consider now a plasma that is heated by some mechanism from $10^6$~K to $10^7$~K in just 1 second, but it takes several minutes for collisions to change the ionization state. In this scenario a plasma of electron temperature $10^7$~K is created with an ionization state (and consequently an emission spectrum) that is characteristic of a $10^6$~K plasma in equilibrium. The time derivative in Equation~\ref{eqn20sb} cannot now be neglected (however, the bulk velocity will be neglected from the total derivative in the following treatment) and a non-equilibrium ionization state arises. In order to determine whether non-equilibrium ionization is important in a particular physical scenario of interest, the equilibration timescale of the ionization state at the new temperature can be estimated from Equation~\ref{eqn20sb}. If it is significantly greater than the timescale of the temperature change itself then non-equilibrium ionization will be important.

\begin{table*}
\caption{Population fractions, and ionization and recombination rate coefficients for the series of ions Fe~XIV,~XV,~XVI. These data are based on the ionization / recombination rate coefficients from / to a given ion provided by \cite{Mazzotta1998,Dere2007}. The rate coefficients are in units of [cm$^3$~s$^{-1}$].}
\begin{tabular*}{\textwidth}{@{\extracolsep{\fill}}c c c c}
\hline
Charge state&Population fraction&ionization rate&Recombination rate\\
    &$10^6$~K&$10^6$~K~~~~$2.5\times10^6$~K&$10^6$~K~~~~$2.5\times10^6$~K\\
\hline
Fe~XIV&$4.60\times10^{-4}$&$4.13\times10^{-12}$~~~~$1.13\times10^{-10}$&$1.35\times10^{-10}$~~~~$5.16\times10^{-11}$\\
Fe~XV &$1.41\times10^{-5}$&$1.04\times10^{-12}$~~~~$6.07\times10^{-11}$&$1.04\times10^{-10}$~~~~$5.09\times10^{-11}$\\
Fe~XVI&$1.40\times10^{-7}$&$3.78\times10^{-13}$~~~~$3.60\times10^{-11}$&$9.66\times10^{-12}$~~~~$2.39\times10^{-11}$\\
\hline
\end{tabular*}
\label{table1}
\end{table*}

Taking a somewhat less extreme example, suppose that a plasma is heated from $10^6$~K to $2.5\times10^6$~K essentially instantaneously. The equilibrium population of Fe~XV reaches its maximum at $2.5\times10^6$~K and so how long does it take to equilibrate in this scenario? Based on the data provided in Table~\ref{table1} we can write:

\begin{eqnarray}
\mbox{Rate of loss of Fe~X} & = & n \left[ -I_i(T+\Delta T)X_i(T) - R_{i-1}(T+\Delta T)X_i(T) \right] \nonumber \\
& = & n \times 1.41\times10^{-5} \times \left( -6.07\times10^{-11} - 5.16\times10^{-11} \right) \nonumber \\
& = & -n \times 1.58\times10^{-15} \mbox{~~~~[s$^{-1}$]};
\label{eqn27sb}
\end{eqnarray}

\begin{eqnarray}
\mbox{Rate of gain of Fe~X} & = & n \left[ I_{i-1}(T+\Delta T)X_{i-1}(T) + R_i(T+\Delta T)X_{i+1}(T) \right] \nonumber \\
& = & n \times \left( 1.13\times10^{-10} \times 4.60\times10^{-4} + 5.09\times10^{-11} \times 1.40\times10^{-7} \right) \nonumber \\
& = & n \times 5.20\times10^{-14} \mbox{~~~~[s$^{-1}$]};
\label{eqn28sb}
\end{eqnarray}

\begin{eqnarray}
\mbox{Net rate of change of Fe~X} & = & n \left( - 1.58\times10^{-15} + 5.20\times10^{-14} \right) \nonumber \\
& = & n \times 5.04\times10^{-14} \mbox{~~~~[s$^{-1}$]}.
\label{eqn29sb}
\end{eqnarray}

\noindent For an electron density characteristic of the solar corona $n=10^9$~cm$^{-3}$ then the equilibration timescale is given by:

\begin{equation}
\tau = \frac{1}{10^9 \times 5.04\times10^{-14}} \approx 20,000 \mbox{~~~~[s]}.
\label{eqn30sb}
\end{equation}

\begin{figure}
\centering
\includegraphics[width=1.0\textwidth]{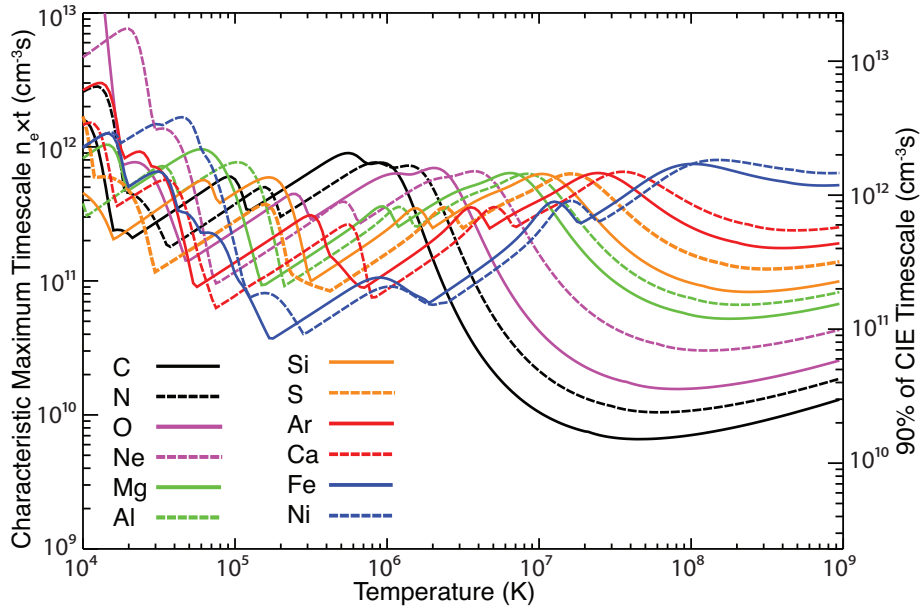}
\caption{\cite{Smith2010}: the left axis measures the density-weighted timescale [cm$^{-3}$~s] for several abundant elements to achieve one e-folding toward ionization equilibrium in a constant temperature plasma; the right axis measures the density-weighted timescale for the elements to reach within 10\% of their equilibrium population.}
\label{fig1sb}
\end{figure}

\noindent If the plasma temperature is changed effectively instantaneously from $10^6$~K to $2.5\times10^6$~K and then held constant at the new temperature, then the population of Fe~XV will approach equilibrium on an e-folding timescale of 20,000~s. In consequence, heating on timescale much shorter than 20,000~s will give rise to a non-equilibrium ionization state; for example, the population of Fe~XV is guaranteed to be out of equilibrium if heating in the solar corona is impulsive (of duration shorter than the characteristic cooling timescale). Heating on timescales significantly longer than 20,000~s allows the ionization state to evolve in equilibrium with the electron temperature. We note that even coronal densities of order $10^{12}$ to $10^{13}$~cm$^{-3}$ may not be sufficient to maintain the ion population close to equilibrium during particularly explosive heating such as occurs during solar flares. \textbf{The estimate of the timescale provided by Equation~\ref{eqn30sb} should be regarded as an absolute upper-limit. The intermediate population fractions of Fe~XV and its neighbouring charge states as the system equilibrates are not accounted for in the approximation. In essence, the rate of change of the population fraction is proportional to the magnitude of the population itself and it will therefore equilibrate more rapidly as it grows.} Figure~\ref{fig1sb} is from \cite{Smith2010} and shows the characteristic equilibration e-folding time-scales for a number of astrophysically abundant elements. \cite{Bradshaw2009} presents a freely available numerical code that solves the time-dependent ionization equations for all elements up to nickel (Z=28), given any tabulated electron temperature and density evolution as a function of time (the time-steps need not be uniform). We describe specific examples of scenarios in which non-equilibrium ionization might arise in Section~\ref{signatures}.

\section{Non-Maxwellian Electron Distributions}
\label{NMED}

Temperature changes on timescales much shorter than those on which ionization and recombination can change the charge state of the plasma are not the only way in which the ionization state may be different than expected for a given temperature. If the electron distribution function is driven away from Maxwellian with the addition of a significant population to the high-energy tail of the distribution, then ions of greater charge can be created at some temperature that is lower than the temperature at which they arise in equilibrium. Non-Maxwellian distributions can arise in several ways; for example, in an astrophysical context they may be expected to occur in circumstances where a region of very hot plasma is separated from a much cooler region by a steep temperature gradient, as is the situation in the solar atmosphere. Collisionless electrons may then stream from the hot, less-dense plasma down the temperature gradient into the cooler, denser plasma driving the tails of the electron distributions in these regions away from Maxwellian.  Non-Maxwellian distributions can also be induced when a beam of particles is accelerated by some mechanism, such as magnetic reconnection, and interacts with the background plasma. A non-Maxwellian distribution can be created in laboratory plasma by laser-heating.

Departures of the electron population from Maxwellian distributions have implications for several properties of the plasma, among them the excitation and ionization states (and, consequently the spectral emission) of its component ions and the transport of energy by thermal conduction. It is therefore highly desirable to take account of these effects in numerical modeling studies, but this is an extremely difficult task to achieve self-consistently. There are two general approaches: (1) carry out particle-in-cell type calculations where the distribution functions can be obtained directly; (2) carry out calculations based on the fluid equations derived by taking successive moments of the underlying distribution function. The first approach is discussed elsewhere in this volume \textbf{(REFERENCES TO ARTICLES IN THIS VOLUME TO BE ADDED)}. The difficulty of the second approach is that solutions to the fluid equations can quickly become inconsistent with the assumptions on which their derivation is based. For example, in the case of steep temperature gradients the mean-free-path of even thermal electrons can approach (and exceed!) the temperature scale length and then the plasma cannot be considered collisional on the characteristic scale length of the fluid. However, the validity of the fluid equations is contingent on the collisionality of the system on the relevant spatial scale. One advantage of the second approach over the first is that plasma systems can be modelled across a much larger range of spatial scales. In particle-type codes one is generally confined to studying phenomena on a particular scale, such as the width of a conduction front or a shock, or the scale of the diffusion region in reconnection \textbf{(REFERENCES TO ARTICLES IN THIS VOLUME TO BE ADDED)}. In fluid codes, the solution can range across many spatial scales from a few meters to hundreds of thousands of kilometers (e.g. in the case of the Sun's atmosphere). The challenge is to develop a method by which physical phenomena on particle scales can be self-consistently included, when needed, in a code that operates predominantly on fluid scales. This requires the distribution function to be calculated from a suitable kinetic equation in tandem with the time-advancement of the system of fluid equations in order that corrections can be made to the fluid variables.

\subsection{Kinetic Equations}
\label{kinetic}

The key to calculating the distribution function in a collisional or weakly-collisional plasma is the manner in which the collision term of the Boltzmann equation is treated. One of two approaches is usually adopted. The first is to handle collisions via a phenomenological term based on the expectation that the plasma particles will ultimately relax into a collisional~/~Maxwellian distribution on some timescale that depends on the degree of collisionality of the plasma. This was first suggested by \cite{Bhatnagar1954} and kinetic equations of this form are commonly referred to as BGK equations (based on the initials of the authors of that paper). 

\begin{equation}
\left( \frac{\partial f_s}{\partial t} \right)_{\mbox{collisions}} = \nu_{ss} \left( F_s - f_s \right) + \nu_{ss'} \left( F_{ss'} - f_s \right),
\label{eqn31sb}
\end{equation}

\noindent where $s,s'$ denote the particle species (e.g. electrons and protons), $\nu_{ss}, \nu_{ss'}$ are the species and inter-species collision frequencies, and $F$ denotes a Maxwellian distribution determined by the local properties of the plasma (e.g. temperature, density and bulk flow). \cite{Morse1963} studied the energy and momentum exchange between non-equipartition gases in the cases of Maxwell, Coulomb and hard sphere interactions, and \cite{Morse1964} showed how to choose free parameters for the cross-collision terms in BGK-type models to conserve density, momentum and energy. This work was limited by the underlying assumption that $n_e / \tau_{ei} = n_i / \tau_{ie}$, which for a fully-ionized hydrogen plasma ($n_e = n_i$) implies that electrons and ions are equally affected by their mutual collisions ($\tau_{ei} = \tau_{ie}$) when in reality they relax on a timescale longer by the square root of the mass ratio ($\tau_{ie} = \sqrt{m_i/m_e} \tau_{ei}$). \cite{Greene1973} then developed a simple improvement for BGK-type models of electron-ion collisions to produce the correct relation between the time scales of ion-electron momentum exchange and ion thermalisation. This work showed how to obtain the correct timescale ordering and how to choose the correct parameters for the Maxwellians in the cross-collision terms to conserve density, momentum and energy.

The second approach to handling collisions is to assume that changes in the velocities of charged particles are due to the cumulative effect of long-range encounters via inverse square forces \citep[e.g.][]{Landau1936}. The collision integral can then be written \citep[e.g.][]{Ljepojevic1990b}:

\begin{equation}
\left( \frac{\partial f_s}{\partial t} \right)_{\mbox{collisions}} = - \Sigma_i \frac{\partial}{\partial v_s^i} \left(f_s \left<\Delta v_s^i \right> \right)\ + \frac{1}{2} \Sigma_{i,j} \frac{\partial^2}{\partial v_s^i \partial v_s^j} \left(f_s \left<\Delta v_s^i \Delta v_s^j \right> \right),
\label{eqn32sb}
\end{equation}

\noindent where

\begin{equation}
\left< \Delta v_s^i \right> =  \Sigma_{s'} \int f_{s'}\left(\mathbf{v_{s'}'} \right) \int_{\theta_{min}}^{\theta_{max}} \sigma_{ss'}(g,\Omega) g \Delta v_s^i d^2\Omega d^3 \mathbf{v_{s'}'}
\label{eqn33sb}
\end{equation}

\noindent and

\begin{equation}
\left< \Delta v_s^i  \Delta v_s^j \right> =  \Sigma_{s'} \int f_{s'}\left(\mathbf{v_{s'}'} \right) \int_{\theta_{min}}^{\theta_{max}} \sigma_{ss'}(g,\Omega) g \Delta v_s^i \Delta v_s^j d^2\Omega d^3 \mathbf{v_{s'}'}.
\label{eqn34sb}
\end{equation}

\noindent The terms of Equations~\ref{eqn32sb} to \ref{eqn34sb} are described in detail in Section~2(a) of \cite{Ljepojevic1990b}. Equations of the form of \ref{eqn32sb} are commonly referred to as Fokker-Planck (FP) equations. \cite{Cohen1950} adopted a method of approximating the distribution function by representing it as a Maxwellian plus a small perturbation to calculate the electrical conductivity of a gas. Their approach is valid in the presence of weak spatial gradients and weak electromagnetic fields. The distribution function then takes the form $f_s=f_0 + f_1$ where $f_0 = F_s$ and

\begin{equation}
f_1 = F_s D(v_s) \mu,
\label{eqn35sb}
\end{equation}

\noindent where $\mu$ is the cosine of the pitch angle. $D$ is a function of the particle speed found by substituting $f_s = f_0 + f_1$ for $f_s$ in the Boltzmann equation, linearising the collisional integral in $f_1$, and solving the integro-differential equation. \cite{Cohen1950} neglected electron-electron interactions from the collision operator but \cite{Spitzer1953}, in what is now considered the `classical' treatment, followed the same approach and included electron-electron interactions in their collision operator. They also extended the solutions to completely ionized gases with different mean nuclear charges and calculate the electrical and thermal conductivities of the gas. The solution to the integro-differential equation in the classical treatment has the form (for electrons)

\begin{equation}
f_e = F_e \left( 1 - \lambda_0 \left[ \frac{ZD_E}{A} \left( \frac{eE}{k_B T_e} + \frac{1}{P_e} \frac{\partial P_e}{\partial s} \right) - 2 \frac{ZD_T}{B} \frac{1}{T_e} \frac{\partial T_e}{\partial s} \right] \mu \right).
\label{eqn36sb}
\end{equation}

\noindent The quantites $ZD_E/A$ and $ZD_T/B$ are tabulated in \cite{Spitzer1953} as functions of the electron speed normalised to the thermal speed and $\lambda_0$ is the mean-free-path of thermal electrons. Above a particular speed $v_{\mbox{crit}}$ the calculated value of $f_1$ becomes comparable to $f_0$ and the assumptions made to derive Equation~\ref{eqn36sb} are invalidated. The approximation of \cite{Spitzer1953} is only valid in the low-velocity regime $v < v_{\mbox{crit}}$ and the upper limit of the regime depends on the strength of the electric field $E$, and the temperature and pressure gradients.

\cite{Rosenbluth1957} derived the Fokker-Planck equation for arbitrary distribution functions in the case where two-body interactions are governed by a force that obeys the inverse square law. The coefficients $\Delta v$ and $\Delta v \Delta v$ in the Fokker-Planck operator were written in terms of two fundamental integrals~/~potentials that depend on the distribution function of the background particles (including those of the same species). Expanding the distribution function as a set of Legendre functions of the pitch angle, the Fokker-Planck equation is cast into the form of an infinite set of one-dimensional, coupled non-linear integro-differential equations. Approximating the distribution function by a finite series, the Fokker-Planck equations can be solved numerically. Keeping one term of the series corresponds to the approximate solution of \cite{Chandrasekhar1943} and keeping two terms yields the solution of \cite{Cohen1950}. \cite{Rosenbluth1957} showed that

\begin{equation}
\left( \frac{\partial f_s}{\partial t} \right)_{\mbox{collisions}} = - \Gamma \frac{\partial}{\partial v_s^i} \left( f \frac{\partial \mathbf{H}}{\partial v_s^i} \right) + \frac{1}{2} \frac{\partial^2}{\partial v_s^i \partial v_s^j} \left( f \frac{\partial^2}{\partial v_s^i \partial v_s^j} \mathbf{G} \right),
\label{eqn37sb}
\end{equation}

\noindent with the Rosenbluth potentials

\begin{eqnarray}
\mathbf{H(v_s)} = \Sigma_{s'} \frac{m_s + m_{s'}}{m_{s'}} \int d^3 v_{s'}' f_{s'}\left(\mathbf{v_{s'}'}\right) \left|v_s - v_{s'}'\right|^{-1} \nonumber \\
\mathbf{G(v_s)} = \Sigma_{s'} \int d^3 v_{s'}' f_{s'}\left(\mathbf{v_{s'}'}\right) \left|v_s - v_{s'}'\right| \nonumber \\
\mbox{where~}\Gamma = \frac{4\pi (Z_se)^2 (Z_se)^2 \ln \Lambda_{ss}}{m_s^2}.
\label{eqn38sb}
\end{eqnarray}

\cite{Ljepojevic1990b} presented a step-by-step description of a method for calculating the distribution function in the presence of strong gradients. In their method the low-velocity part of the distribution is given by solution of \cite{Spitzer1953}. The high-velocity tail of the distribution function is given as a solution to the high-velocity form of the Fokker-Planck equation which is derived from Equations~\ref{eqn37sb} and \ref{eqn38sb} by neglecting the interaction between the high-velocity particles themselves and considering only their interaction with the low-velocity (near Maxwellian) part of the distribution function. One may then derive a linearised form of the Boltzmann equation with the Fokker-Planck collision operator that applies to high-velocity particles. For electrons:

\begin{equation}
\left( \frac{\partial f_e}{\partial t} \right)_{\mbox{collisions}} = \frac{1}{v_e^2} \frac{\partial}{\partial v_e} \left[ v_e^2 \nu(v_e) \left( \frac{k_B T_e}{m_e} \frac{\partial f_e}{\partial v_e} + v_e f_e \right) \right] - \nu(v_e) \frac{\partial}{\partial \mu} \left[ \left( 1-\mu^2\right) \frac{\partial f_e}{\partial \mu} \right].
\label{eqn39sb}
\end{equation}

\noindent The full derivation of Equation~\ref{eqn39sb} is given by \cite{Ljepojevic1990b} on pages 73 to 88 of their article. They also describe in detail a numerical treatment for its solution following non-dimensionalisation and transformation into a form more convenient for numerical work. The solutions in the low-velocity and high-velocity regime are combined, subject to suitable matching conditions (e.g. smoothness), in a region of the velocity space where both methods are approximately valid; two thermal speeds was found to be the optimal value.

Given the significant complexities that are involved in working with the Fokker-Planck equation it is tempting to revert to BGK-type approximations of the collision operator. However, one must be careful. \cite{Livi1986} compared the collisional relaxation of a double-beam and a bi-Maxwellian distribution function for a Fokker-Planck and a BGK collision operator. They found that moments of the distribution function up to and including temperature (the 2nd moment) were in good agreement between the two schemes when the frictional energy-loss rate was used as the effective collision frequency in the BGK operator, but that the heat flux (the 3rd moment) exhibited differences due to its sensitivity to the shape~/~skew of the distribution function, which enters the Fokker-Planck operator via the second derivative of the distribution function w.r.t. velocity. \cite{Ljepojevic1988} calculated contributions to the heat flux in a solar flare atmosphere from the tail of the distribution function using the high-velocity form of the Fokker-Planck equation \citep{Ljepojevic1990b} and compared the results with a BGK-type calculation. They found that the BGK technique can estimate contributions from the high-energy tail to the heat flux to within order of magnitude.

As computers became more powerful, detailed numerical treatments of the Fokker-Planck equation became feasible. \cite{Shoub1983} provides a detailed discussion of the break-down of the \cite{Spitzer1953} calculation of the electron distribution function and describes an approach to deriving and then solving numerically the high-velocity form of the Landau-Fokker-Planck equation. Implications of the break-down of the local Maxwellian approximation are discussed for: energy balance in the upper chromosphere and low TR; the He resonance line spectrum; the Schmahl-Orrall observation of continuum absorption by neutral H, and the origin of the 20,000~K temperature plateau. However, \cite{Shoub1983} was unable to say anything quantitative about the heat flux since the kinetic equation was only solved to six thermal speeds. Had \cite{Shoub1983} applied the same transformation following non-dimensionalisation as employed by \cite{Ljepojevic1990b} then it would have been possible to significantly extend the calculation in velocity space. \cite{Ljepojevic1990a} used the approach described in \cite{Ljepojevic1990b} to show that distribution functions are near Maxwellian in the commonly used FAL \citep{Fontenla1993} models of the photosphere to mid-TR and so the models are valid in their given form. \cite{MacNeice1991} applied the same approach to the transition region of a flaring loop and found a substantial enhancement in the tail populations throughout that region of the atmosphere. We discuss the results of some attempts to apply these calculations of distribution functions to fluid models in order to take account of the consequences of non-Maxwellian distributions in the following Section.

\subsection{Heat Flux~/~Transport}
\label{heatflux}

The fluid equations are derived by taking successive moments of the Boltzmann equation when it is written in terms of distribution functions that exhibit only small deviations from a fully-relaxed Maxwellian distribution. Since the statistical treatment of a particle ensemble in terms of a fluid is valid only in this collisional limit, then only small deviations can be tolerated. In general, departures from Maxwellian are treated as a perturbation and the distribution is expanded in terms of some parameter that should remain small, such as the ratio of the electron mean-free-path to the temperature scale length (the Knudsen number, $Kn$), in order to derive non-linear terms of the fluid equations such as the heat flux.

The transport of heat by thermal conduction is the dominant transport process in hot but tenuous astrophysical plasmas. It determines the temperature and thus the density structure, via the temperature-dependent scale length, in the solar atmosphere (for example) and so it is important to handle it as accurately as possible. The most commonly used form for the heat flux is that given by \cite{Spitzer1953}, valid in the limit of weak gradients and weak electric fields:

\begin{equation}
F_c = - \kappa \nabla T,
\label{eqn40sb}
\end{equation}

\noindent where the conductivity $\kappa = \kappa_0 T^{5/2}$ (for a fully ionized hydrogen plasma) and the constant is the quantity calculated by Spitzer ($\kappa_0 \approx 10^{-6}$). Despite its strong non-linearity the form represented by Equation~\ref{eqn40sb} is convenient to implement in fluid-based numerical codes, but cannot guarantee an accurate representation of the heat flux if it is used indiscriminately. Experimental and numerical results have shown that its range of applicability is actually quite limited. Equation~\ref{eqn40sb} indicates that the heat flux can increase indefinitely provided that the temperature gradient continues to steepen, but eventually a physical limit must be reached when there are no more particles remaining to support the implied heat flux. This is the free-streaming limit, essentially the maximum heat flux that the plasma can sustain, and may be estimated by assuming that the majority of the particles (e.g. electrons) stream down the temperature gradient at the thermal speed \citep{Bradshaw2006} (more sophisticated numerical treatments indicate the free-streaming limit is about 1/6 of this value). At the very least, then, a limiter should be deployed in any numerical model that uses Equation~\ref{eqn40sb}, in order to constrain the heat flux to physically justifiable values.

There have been a number of efforts to derive systems of fluid equations that take account of stronger departures from Maxwellian to be implemented in numerical models. \cite{Campbell1984} found a solution to the Boltzmann equation that extends the Chapman-Enskog approximation to large temperature gradients and electric fields, to calculate electron transport in a fully ionized gas. The collision term was written in the form of a collisional relaxation with a velocity-dependent relaxation time defined in terms of the scattering length. The distribution function was assumed to be separable with the angular dependence represented by a slowly varying function. Calculating the moments of this distribution function led to correction factors to the classical \citep{Spitzer1953} transport coefficients as a function of the temperature gradient scale-length and an inherently flux-limited heat flow. \cite{Killie2004} derived a complete set of fluid equations for fully ionized gases that improve the treatment of Coulomb collisions by taking into account the shape of the distribution function to better calculate the heat flux and the thermal force. They chose an analytical velocity distribution function with a Maxwellian core plus a high-velocity correction term proportional to $v^3$, and obtained transport equations by inserting their choice of distribution function into the Boltzmann equation with a Fokker-Planck collision operator. \cite{Chiuderi2011} derived a set of two-fluid equations applicable to weakly collisional plasmas by using a relaxation approach to the collision operator and selecting `mixed' Maxwellian distributions for the two interacting species that conserve momentum and energy. The collisional term in their treatment depends on an `average' or `representative' collisional timescale that is velocity-independent.

\cite{Gray1980} described the results of experiments in which the ratio of the electron mean-free-path to the temperature scale-length was found to be about 0.5 and enhanced low-frequency turbulence was observed. They used a numerical simulation of the experimental set-up to show that ratios of 0.5 implied a heat flux limited to less than 5\% of the free-streaming limit in the hot part of the plasma. They also found $T_e/T_i$ in the same region sufficient to excite heat-flux driven ion acoustic turbulence, thus explaining the low frequency turbulence observed in the experiment. The observed level of turbulence in the experiment was enough to account for the predicted low thermal conductivity in the numerical model, which was due to electron scattering from interactions with the ion acoustic turbulence. \cite{Bell1981} and \cite{Matte1982} studied electron heat transport down steep temperature gradients in laser-induced plasmas by numerically solving the Fokker-Planck equation. The heat flux was found to be substantially smaller than that predicted by the classical theory or the free-streaming value when the mean-free-path reached a fraction of only one-hundredth of the temperature scale length ($Kn=10^{-2}$). \cite{Shoub1983} found significant deviations from Maxwellian in the tail of the distribution for $Kn=10^{-3}$, but was unable to provide a quantitative estimate of the heat flux. \cite{Owocki1986} used a BGK-type method to calculate the electron distribution function in the solar transition region to study the effect of a high-energy tail on the heat transport and collisional excitation and ionization rates. For the case studied they found that non-classical transport does not significantly alter the excitation or ionization state of ions with emission lines that form predominantly in the lower transition region (with excitation energies in the range 10 eV, because electrons with these low energies thermalise quickly), but the non-classical heat flux in this region does depend sensitively on the temperature gradient in the upper transition region.

In the case of pronounced departures from Maxwellian distributions it is clear that correction factors and localised approaches to calculating the distribution function, and hence the heat flux, are not sufficient. For example, contributions to local quantities from non-local sources may lead to strong departures from local Maxwellian distributions. Such kinetic behaviour is inherently incompatible with the fluid approximation in which it is assumed that the properties of the plasma can be determined entirely locally (e.g. the heat flux as a function of temperature and the temperature gradient). Since it is generally not feasible to solve a kinetic equation (certainly not a time-dependent form) in tandem with the fluid equations to correct for the consequences of kinetic behaviour, then the challenge is to find an alternative; e.g. a computationally tractable approach that can be implemented in an otherwise fluid-based treatment, and that permits one to account for purely kinetic effects (e.g. non-local influences) on quantities such as the heat flux in regions where the Knudsen number grows large. One such approach is to adopt a delocalisation formula.

\cite{Luciani1983} found a delocalisation formula for the heat flux, using a set of solutions to the Fokker-Planck equation. Delocalisation formulae are based on delocalisation kernels that operate on calculations of the heat flux made using Equation~\ref{eqn40sb}. The kernel essentially acts to `smear' the classical heat flux out over the computational domain in a manner that mimics the spatial profile of the heat flux that would be found from a full Fokker-Planck calculation. The formula presented by \cite{Luciani1983} has the form:

\begin{equation}
F_c(s) = \int w(s,s') F_{SH}(s') ds'
\label{eqn41sb}
\end{equation}

\noindent and the delocalisation kernel is

\begin{equation}
w(s,s') = \frac{1}{2\lambda(s')} \exp \left[- \left| \int_{s'}^s \frac{n(s'')}{\lambda(s')n(s')} ds''  \right| \right].
\label{eqn42sb}
\end{equation}

The quantity $\lambda$ is an effective range for the electrons, related to the mean-free-path. In the limit of shallow temperature and density gradients the kernel $w$ behaves like a $\delta$-function, where $\int w(s,s') ds' = 1$ and Equation~\ref{eqn41sb} reduces to $F_c = F_{SH}$. Despite the double integration, the delocalisation formula is straightforward to efficiently implement in a fluid code to replace the heat flux in the form of Equation~\ref{eqn40sb}. \cite{Luciani1985} found an analytical justification for the delocalisation formula and \cite{Bendib1988} developed an improvement that takes the presence of an electric field into account. More recently, \cite{AlouaniBibi2004} studied non-local electron heat transport using a number of different approximations to the Rosenbluth potentials in the Fokker-Planck equation to find delocalisation kernels for non-local heat flux formulae to be used in fluid codes.

A number of authors have compared the different approaches to calculating the heat flux and have implemented them in numerical models in order to apply them to particular problems in which non-Maxwellian electron distributions are expected to arise. \cite{Smith1986} discussed classical \citep{Spitzer1953}, locally limited \citep{Campbell1984} and non-local \citep{Luciani1983,Luciani1985} algorithms for the heat flux and their application to heat transport in the case of the steep temperature gradients (thin conduction fronts) that arise during the impulsive phase of solar flares. \cite{Karpen1987} investigated how these different formulations for the heat flux affect the physical characteristics of the corona, transition region and chromosphere in numerical models of solar flares. Both sets of authors found that the heat flux in the hot part of the plasma obtained with the non-local treatment was smaller than the locally limited and classical values, whereas the heat flux in the colder parts of the plasma (e.g. in the transition region and chromospere) was significantly enhanced compared with the locally limited and classical values. In consequence, both flux limiting and delocalisation play an important role in the evolution of the plasma. In the case of flares this leads to a `bottling up' of energy in the corona, allowing it to reach much higher temperatures, and the earlier onset of weaker chromospheric evaporation.

\begin{figure}
\centering
\includegraphics[width=1.0\textwidth]{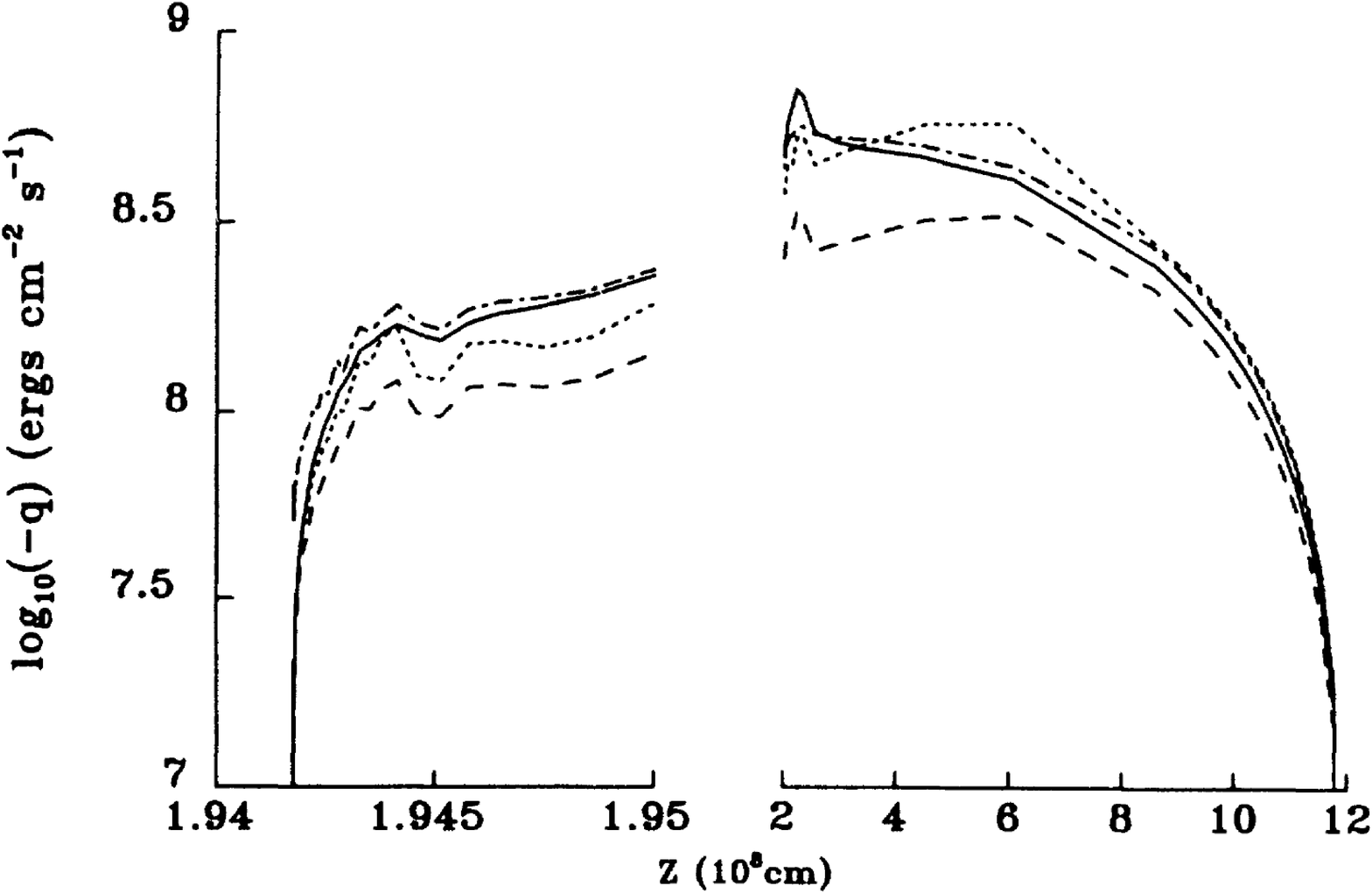}
\caption{The heat flux in a solar active region loop calculated by \cite{Ljepojevic1989,Ljepojevic1990b} (solid line), \cite{Spitzer1953} (dotted line), \cite{Campbell1984} (dashed line), and \citep{Luciani1983,Luciani1985} (dot-dashed line). Image credit: \cite{Ljepojevic1989}.}
\label{fig9}
\end{figure}

\cite{Ljepojevic1989} calculated the heat flux in a solar active region coronal loop from distribution functions obtained using the sophisticated model described in \cite{Ljepojevic1990b}, and compared it with the heat flux given by the classical treatment of \cite{Spitzer1953}, the correction coefficients to the classical treatment given by \cite{Campbell1984} and the heat flux given by the delocalisation formula of \cite{Luciani1983,Luciani1985} (Figure~\ref{fig9}). They concluded that the classical treatment failed completely in the lower corona, predicting a strong heat flux flowing down the temperature gradient when the kinetic equation yielded heat flux flowing up the temperature gradient, and the possibility that the role of the heat flux could be misinterpreted in the energy balance of the corona. \cite{Landi2001} also found that the heat flux can flow up the temperature gradient in the case of supra-thermal tails characterised by $\kappa$ distributions with $\kappa < 5$. Encouragingly, \cite{Ljepojevic1989} did find relatively good agreement between the delocalisation formula and the more sophisticated kinetic calculation. \cite{West2008} implemented the delocalisation formula in the HYDRAD \citep[e.g.][]{Bradshaw2003b,Bradshaw2012} code to investigate the lifetime of hot, nanoflare-heated plasma in the solar corona. The aim of this work was to determine whether the bottling up of energy in the corona due to the severe heat flux limiting that arises in the limit of large Knudsen numbers provided sufficient time for the ionization state to equilibrate following rapid heating. \cite{AlouaniBibi2002} developed a non-local model of electron heat flow in laser-heated plasmas, taking into account super-Gaussian deformation of the electron distribution function. \cite{AlouaniBibi2003} derived an analytical description of electron-ion energy exchange by Coulomb collisions in the presence of super-Gaussian electron distributions, and found the ratio $T_i / T_e$ at which the collisional electron-ion energy exchange cancels increases from 1 in a Maxwellian plasma to 1.98 in a super-Gaussian plasma.

\subsection{Excitation, Ionization and Radiation}
\label{eir}

The specific nature of the local distribution function can have an important effect on the rate of excitation and ionization via collisions. Excitation is generally a consequence of interactions between ions and electrons in the bulk of the distribution, but ionization is particularly sensitive to the tail population. The stronger heat fluxes at the base of steep temperature gradients found in a number of the studies described in Section~\ref{heatflux} imply enhanced tail populations of streaming electrons, which can feasibly alter the ionization state such that it can no longer be considered a strong function of the local temperature (and, to a lesser extent, the local density) alone. Collisional excitation rate coefficients can be calculated by substituting a suitable distribution function into Equation~\ref{eqn10sb} and the ionization rate can be calculated by inserting the appropriate ionization cross-section (usually pertaining to the ground-state) in place of $Q_{i,j}$. Investigations of the effect of non-Maxwellian distributions on the ionization state have proceeded along two general lines: (a) calculate the distribution function by solving some simplified form of the Boltzmann equation (e.g. BGK, Fokker-Planck); or (b) choose an analytical form for the distribution function with the properties of a Maxwellian at low-velocities~/~energies, but which permits an enhanced tail population where the degree of enhancement can be controlled by a single parameter. A popular generalisation of the Maxwellian distribution that fulfils these requirements is the $\kappa$-distribution:

\begin{equation}
f_\kappa(E) = A_\kappa \left( \frac{m}{2\pi k_B T} \right)^{3/2} \frac{\sqrt{E}}{\left(1+\frac{E}{\left(\kappa-1.5\right)k_BT}\right)^{\kappa+1}};
\label{eqn43sb}
\end{equation}

\begin{equation}
A_\kappa = \frac{\Gamma\left(\kappa+1\right)}{\Gamma\left(\kappa-0.5\right)\left(\kappa-1.5\right)^{3/2}}.
\label{eqn44sb}
\end{equation}

\noindent The $\kappa$-distribution has the form of a Maxwellian in the limit $\kappa \rightarrow \infty$. The most probable energy of a particle in the distribution is $E_p =(\kappa-1.5)k_BT/\kappa$ and the mean energy of the distribution is $<E>=3k_BT/2$ (i.e. independent of $\kappa$ and the same as the Maxwellian at the same temperature). \cite{Yoon2006} and \cite{Rhee2006} demonstrated that $\kappa$-distributions can be induced by spontaneous scattering (absent in collisional treatments) when electron beams are accelerated by weakly turbulent processes.

\cite{Owocki1982} used $\kappa$-distributions to study the ionization state of gases with non-Maxwellian electron distributions, finding changes from the ionization temperature assuming an underlying Maxwellian distribution of up to a factor of 2. The importance of the high-velocity tail to the ionization state depends on the ratio of the ionization potential to the mean thermal energy of the electrons. \cite{Owocki1982} also found that the high ionization energy required for the O~VIII~$\leftrightarrow$~O~IX transition means that oxygen ionization at solar coronal temperatures is more sensitive to the tail of the distribution than elements of lower ionization energy (such as iron) found within that temperature range. \cite{Dzifcakova2003} calculated excitation, ionization and recombination rates for $\kappa$-distributions for a range of values of the parameter $\kappa$. They found changes in the level populations and the relative ion abundances. A synthetic spectrum was calculated which showed that some C~III, C~IV and O~IV lines are sensitive to the shape of the distribution function and their intensities enhanced by a factor $2-6$ in the presence of strongly non-thermal distributions. \cite{Dzifcakova2006} investigated the influence of $\kappa$-distributions in the solar corona on Fe~VIII - XV excitation and ionization, and on the line intensities associated with those transitions. They concluded that it ought to be possible to diagnose the value of $\kappa$ that would best characterise the electron distribution from ratios of Fe~IX 171~\AA, Fe~XII 195~\AA~and Fe~XV 284~\AA~lines provided that the plasma density is known.

\cite{Dzifcakova2008a} calculated non-Maxwellian electron excitation rates for ions of astrophysical interest. They demonstrated a method for extracting collision strengths from the Maxwell-averaged collision strengths (Upsilons) that are provided by the CHIANTI atomic database and then integrated these over the specific non-Maxwellian distribution in order to calculate the corresponding excitation rate. $\kappa$-distributions, employing a range of values of $\kappa$, were used to calculate synthetic spectra for Fe~XV and XVI in the $50-80$~\AA~range for comparison with solar observations. For consistency in the generation of the synthetic spectra, \cite{Dzifcakova2008a} used the equilibrium ionization states of Fe derived for a range of $\kappa$ by \cite{Dzifcakova2002}. However, they found no conclusive evidence for non-Maxwellian distributions in the particular flare dataset that was compared with the synthetic spectra.

\begin{figure}
\centering
\includegraphics[width=1.0\textwidth]{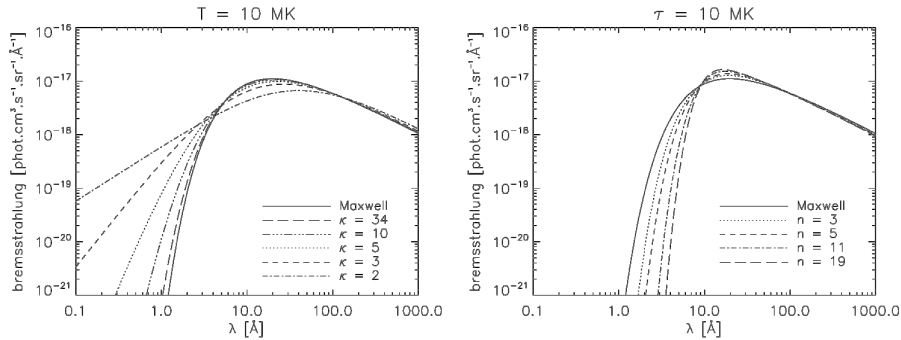}
\caption{The X-ray and UV continuum at 10~MK calculated from $\kappa$ and $n$ distributions representing different strengths of departure from Maxwellian. Image credit: \cite{Dudik2012}.}
\label{fig10}
\end{figure}

\cite{Dudik2011} calculated the bound-bound and free-free radiative losses arising from plasmas with non-Maxwellian electron distribution functions using $\kappa$- and $n$-distributions  \citep[Equations~5 and 6 of][]{Dudik2011}. It was found that changes in the radiative loss function due to non-Maxwellian distributions are greater than errors in the atomic data and errors due to missing contributions from free-bound continuum. While radiative loss functions for $\kappa$ distributions are generally weaker than for Maxwellians, the opposite is true for $n$-distributions. They also found that the contribution from bremsstrahlung changes by only a few percent, except in the extreme case of $\kappa=2$. Following on from this earlier work, \cite{Dudik2012} calculated the X-ray, UV and radio continuum arising from non-Maxwellian distributions using $\kappa$- and $n$-distributions (Figure~\ref{fig10}). They found that at flare temperatures and hard X-ray energies both the bremsstrahlung and the free-bound spectra are dependent on the assumed distribution, and concluded that the low energy part of $\kappa$ distributions can be determined from observations of the continuum.

\section{The Radiative Loss Function}
\label{RLF}

\begin{figure}
\centering
\includegraphics[width=1.0\textwidth]{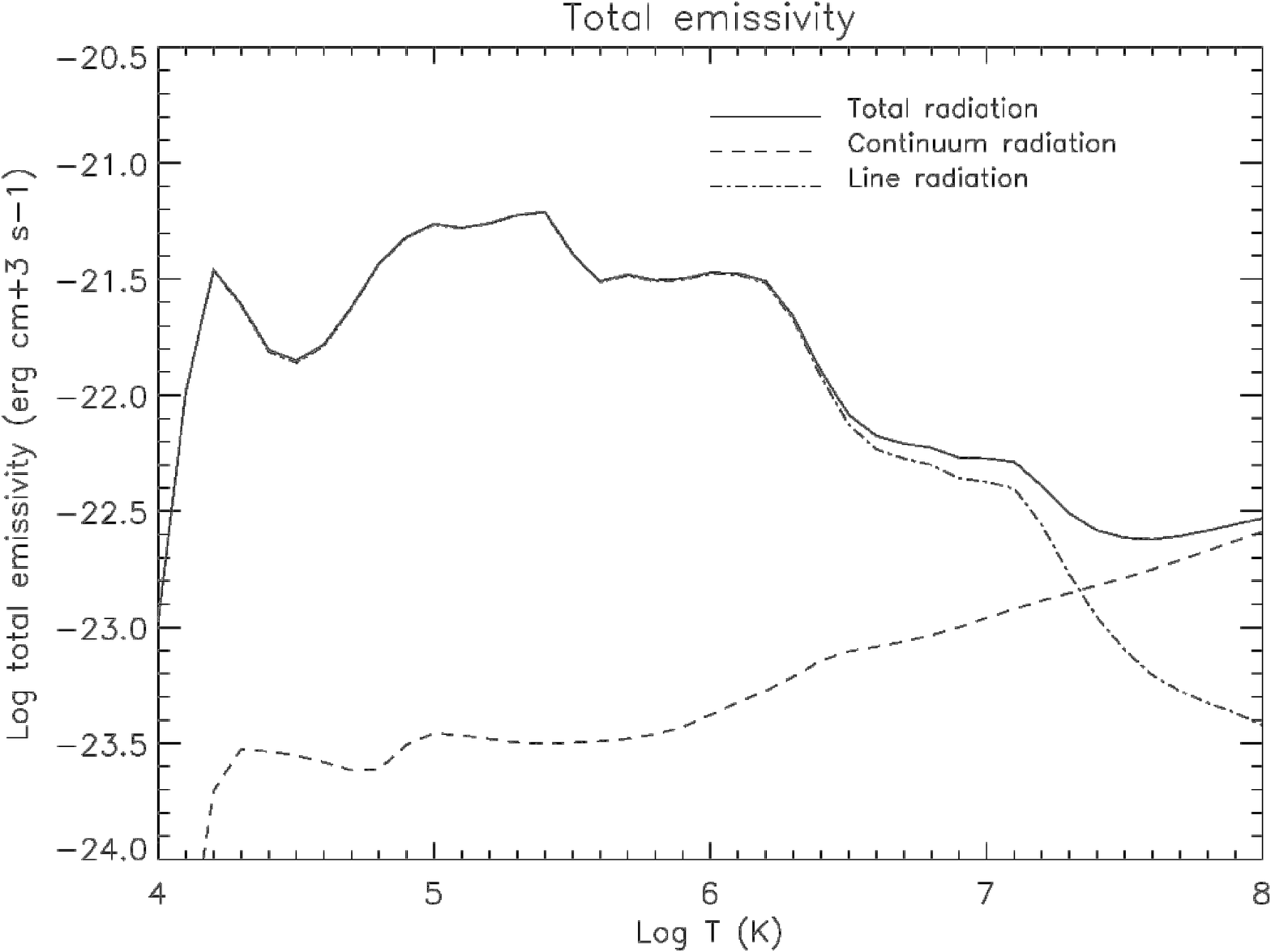}
\caption{The optically-thin radiative loss function. Image credit: \cite{Landi1999}.}
\label{fig11}
\end{figure}

The power per unit volume of plasma emitted by a single spectral line is given by Equation~\ref{eqn7sb}. The total power per unit volume is then the sum total of the power emitted by the many thousands of spectral lines which belong to the chemical elements that comprise the plasma. In the case of optically-thin astrophysical plasmas this quantity can be written in a conveniently compact form:

\begin{equation}
E_R = N_e N_H \Lambda(T_e) \mbox{~~~~[erg~cm$^{-3}$~s$^{-1}$]},
\label{eqn45sb}
\end{equation}

\noindent where $N_H$ is the number density of hydrogen atoms. In a fully ionized hydrogen plasma $N_e=N_H$. $E_R$ is generally referred to as the {\it radiative volumetric loss rate}. $\Lambda(T_e)$ is called the {\it optically-thin radiative loss function} (or the total emissivity of the plasma, as shown in Figure~\ref{fig11}) and it encapsulates a vast amount of atomic data. The radiative loss function depends upon the element abundances, the ionization state of the elements, and transition energies and probabilities. These must be determined for hundreds of ions and in many cases thousands of spectral lines per ion, in order that accurate radiative loss functions can be calculated. As atomic data is improved and updated then so must the radiative loss function. The most convenient way to keep abreast of developments is to use a comprehensive and regularly updated atomic database, such as Chianti \citep[][: https://www.chiantidatabase.org]{Dere1997,Landi2012}, which contains carefully assessed atomic data and the appropriate functionality for calculating spectra and $\Lambda$. A number of assumptions concerning the properties of the plasma, such as the nature electron distribution function and the time-dependence of the ionization state, are implicit in the most commonly used calculations of $\Lambda$ for astrophysical applications. The limits of these assumptions have been demonstrated in Sections~\ref{ionization} and \ref{NMED} and so the emissivity of individual spectral lines and the radiative loss function must be recalculated, whenever these limits are reached, to take proper account of the microphysical processes that can arise in astrophysical plasmas.

Following Equation~\ref{eqn7sb} the {\it emissivity} of a single transition between energy levels $j$ and $g$ in a particular ion is given by

\begin{equation}
\epsilon_{j,g} = \frac{N\left(X^{+m}\right)}{N(X)} \frac{N(X)}{N(H)} \frac{N(H)}{N_e} C_{g,j}^e \Delta E_{j,g} \mbox{~~~~[erg~cm$^3$~s$^{-1}$]}.
\label{eqn46sb}
\end{equation}

\noindent The {\it ion emissivity} is then obtained by summing over all of the transitions that may occur within the ion:

\begin{equation}
\Lambda_{X_i} = \Sigma_{\lambda} \epsilon_{j,g},
\label{eqn47sb}
\end{equation}

\noindent so that $\Lambda_{X_i}$ is the radiative loss function for the particular ionization state $i$ of element $X$. The radiative loss function for the element can be found by summing over the number of charge states:

\begin{equation}
\Lambda_X = \Sigma_i \Lambda_{X_i}
\label{eqn48sb}
\end{equation}

\noindent and the total radiative loss function is the sum over the number of elements of interest:

\begin{equation}
\Lambda = \Sigma_X \Lambda_X.
\label{eqn49sb}
\end{equation}

\noindent The radiative loss function is strongly dependent on the electron temperature $T_e$ in equilibrium, but it is clear from Equation~\ref{eqn46sb} how important the ionization state (the first factor on the right-hand side) and the collisional excitation rate $C_{g,j}^e$ are to accurately calculating it. When the ionization state exhibits strong departures from equilibrium then the temperature dependence of the radiative loss function can be lifted, the rate of energy loss by radiation, and the intensities of individual spectral lines, may not be characteristic of the actual electron temperature. Furthermore, the ionization state (via the ionization and recombination rate coefficients) and the collisional excitation rate depend on the underlying electron distribution which is generally assumed to be Maxwellian, but circumstances can easily arise astrophysical plasmas when this assumption is certainly not valid (Section~\ref{NMED}).

\section{Signatures and Diagnostics of Non-Equilibrium Processes}
\label{signatures}

Analytical analyses can identify the conditions under which non-equilibrium processes become important to understanding the properties and behaviour of astrophysical plasmas, and numerical models demonstrate that such conditions are commonplace in the optically-thin astrophysical plasma systems that are the focus of a great deal of current research interest. In this Section we consider potential signatures of non-equilibrium processes and the evidence for their manifestation in observational datasets.

\subsection{Non-equilibrium Ionization}
\label{IonNEq}

\cite{Griem1964} discussed the potential for departures from equilibrium of the ionization state in cases where the dynamical evolution of the plasma occurs on timescales that are shorter than those of ionization and recombination. He cited the particular example of transporting the ion population across a strong temperature gradient, as might be the case in the solar transition region. \cite{Joslyn1979a,Joslyn1979b} investigated steady flows across a range of temperature gradients and found that ionization equilibrium in the transition region is an acceptable assumption for iron at flow speeds no greater than 20~km/s, but that carbon and oxygen ion populations can be driven away from equilibrium at flow speeds of only 1~km/s. \cite{RaymondDupree} and \cite{Dupree1979} carried out a similar study related to steady flows in the transition region and also found significant departures of the ion populations from equilibrium. \cite{Borini1982} investigated the ionization state in coronal loops in the temperature range 0.2~MK to 2~MK and showed that considerable deviations from equilibrium ionization could arise in average to low intensity loops characterised by high-speed flows. They reported a pronounced effect for cooler loops, which despite exhibiting lower speed flows were found to have steeper temperature and density gradients than hot loops.

\cite{Noci1989} calculate the number density of carbon ions for a selection of coronal loops models in the case of steady-state, sub-sonic flows (siphon flows) and found departures from equilibrium of the ionization state for flows of only a few km/s at the loop apex and for a factor of 10 slower at the base of the transition region. \cite{Spadaro1990a} calculated the spectral line profiles of carbon ions formed in the transition region that are commonly used in spectroscopic diagnostic studies. They used the number densities of the carbon ions found by \cite{Noci1989} and found predominantly blue-shifted emission lines, which could not be reconciled with observations that show both up- and down-flows in the transition region. The absence of red-shifted emission was attributed to the assumption of spatially uniform heating. \cite{Spadaro1990b} focused on the corona and calculated the emissivities of carbon and oxygen, both in and out of equilibrium, and found substantial differences between the resulting radiative loss functions. In the case of up-flows (down-flows) the radiative losses were generally enhanced (suppressed). One may understand this by considering an ion of relatively low charge state transported into a region of temperature significantly higher than the formation temperature of the ion in equilibrium; the ion will tend to emit more strongly since a greater proportion of the electrons in the bulk of the distribution will have sufficient energy to excite its emission lines. 

\cite{Spadaro1994a,Spadaro1994b} studied the effect of non-equilibrium ion populations on the line intensities of O~VI and H~I ions that originate in solar wind source regions. They calculated the intensity and line profiles for equilibrium and non-equilibrium ionization balance based on a steady flow model, finding significant deviations from equilibrium beyond $3-4$ solar radii for O~VI and beyond 5 solar radii for the Lyman~$\alpha$ emission from H~I. These results are significant for estimates of the solar wind speed that rely on the Doppler-dimming technique, which estimates the speed from variations in the line intensities compared with their expected values in the absence of a steady outflow.

\cite{Spadaro1994c} investigated the signatures that may be observable when non-equilibrium ion populations are present and should be considered when carrying out spectroscopic diagnostics using line ratios. Since non-equilibrium ion populations are displaced from their temperatures of peak abundance in equilibrium, the temperature-dependent Boltzmann factors that appear in the expression for the excitation rate coefficients for the spectral lines are changed considerably, which results in changes to the energy level populations, the line intensities and, consequently, the values of the line ratios. The values of the temperature sensitive line intensities arising from the non-equilibrium C~IV and O~IV-VI populations calculated  by \cite{Noci1989} (carbon) and \cite{Spadaro1990b} (oxygen)  were compared with the same line intensities computed in equilibrium. In the presence of a non-equilibrium ion population the line intensities were found to be reduced for both up-flows and down-flows across the transition region temperature gradient. The C~IV population was found to be the most sensitive to non-equilibrium ionization, with decreases in the line ratio by an order of magnitude in the case of down-flows. In response to discrepancies identified by \cite{Keenan1992} between C~IV line intensities observed during highly dynamic events and theoretical predictions of the same line intensities, \cite{Spadaro1995} used a siphon-flow model and non-equilibrium ion populations to recalculate the predicted line intensities. However, they found only a marginal improvements in the agreement between the observed and predicted intensities when non-equilibrium ionization was accounted for, and conlcuded that the observed intensities could not be reconciled with a sub-sonic, siphon-flow model. \cite{Esser1998} examined the effect on the ionization state of the solar wind when the acceleration process occurs at much lower heights in the solar atmosphere than previously considered, based on flow speeds estimated from chromospheric, transition region and coronal emission lines. These observations yielded flow speeds for O~VI ions that are a factor of $3-4$ greater than indicated by earlier work, which imply the ion populations may depart from equilibrium as they are transported at speed across the steep temperature gradients found in the lower atmosphere. In this case, the use of charge state ratios to estimate equilibrium temperatures is unlikely to be valid. \cite{Esser1998} found outflow models with speeds in the region of $130-230$~km/s to predict charge state ratios consistent with those observed. \cite{Edgar2000} considered the effect of non-equilibrium ionization on the ratio of Ne~VI to Mg~VI lines in the solar transition region, which is used as a diagnostic of the first ionizaton potential (FIP) effect. In the presence of a strong heating or cooling effect the populations of ions of low FIP are enhanced relative to those of higher FIP. They calculated the non-equilibrium populations of these ions for simple flows across the transition region and showed that their spectral line ratios depend on non-equilibrium effects, as well as on the temperature and density.

These investigations into the consequences of non-equilibrium ionization assumed steady-state conditions where only flows may affect the ionization state in the presence of a steep temperature gradient. In general, this is due to the assumption of some form of constant heating that maintains the plasma in a steady-state condition, but this need not be so. There exist mechanisms by which energy can be impulsively released into the plasma on short timescales (e.g. a collisionless shock or magnetic reconnection, \textbf{(REFERENCES TO ARTICLES IN THIS VOLUME TO BE ADDED)}) leading to temperature changes on timescales that are short compared with the ionization time.  Local temperature enhancements can give rise to localised pressure gradients which may in turn drive flows. Consequently, a detailed understanding of the consequences of non-equilibrium ionization requires a treatment of both local, temporal changes in the plasma properties and the fast transport of ions by flows. A shock is perhaps the simplest case, since it drives the plasma from one nearly steady state to another, and if the shock is stong the ionization state can be far from equilibrium. \cite{ma11} used the compression, density and heating determined from optical and radio observations of a CME-driven shock to compute the time-dependent ionization in the post-shock flow, and they found that it matched the observed rise times of emission in the AIA bands.

\cite{Hansteen1993} presented a numerical model that solved the time-dependent ion population equations in tandem with the hydrodynamic equations, taking account of departures from ionization equilibrium on the radiative losses for ions formed below 0.3~MK. The model was used to study the dynamic response of a coronal loop to energy released impulsively near the apex. It was found that the line shifts predicted for C~IV, O~IV and O~VI by the model were consistent with the persistent red-shifts observed in transition region lines \citep[e.g.][]{Brekke1997}. The amplitude of the predicted line shift was shown to depend on the ionization timescale of the emitting ion. It was also found that the radiative losses could change by a factor of 2 due to the influence of flows and waves on the ion population. \cite{Teriaca1999a} noted the presence of blue-shifts at temperatures characteristic of the transition region in the quiet Sun and in active regions and \cite{Teriaca1999b} suggested that impulsive heating localised in the transition region at the temperature of peak O~VI abundance in equiibrium (0.3~MK) might account for the presence of red-shifts and blue-shifts. The heating (whether located at the loop apex or in the transition region) generates compression waves and by including the partial reflection of the downward propagating wave from the chromosphere, and allowing for non-equilibrium ionization, reasonable agreement was found between the observed Doppler-shifts and those predicted by the numerical model (red-shifts in the cooler C~IV lines and blue-shifts in the warmer O~VI lines). \cite{Doyle2002} found that the higher the temperature at which a heating event occurs then the greater the delay in the response from the mid-transition region lines in terms of changes in the Doppler-shift.

Bradshaw \& Mason (2003) studied the response of the plasma and the ionization state to a small-scale, impulsive energy release at the apex of a coronal loop, characteristic of nanoflare heating and solved the ion population equations for the 15 most abundant elements of the solar atmosphere (including C, O, Ne, Mg, Si and Fe). The ionization state was used to calculate the radiative loss function in the energy equation, thereby coupling the energy balance with the ionization state. They concluded that broad~/~narrow-band imaging instruments can miss small-scale heating events entirely due to the weak sensitivity of the non-equiibrium emissivity to the changing temperature compared to the emissivity for equilibrium ionization, which fell by a factor of up to 5. The non-equilibrium emission remained relatively steady throughout the heating event, despite a factor 2 change in the temperature on a timescale of 30~s. In order to diagnose non-equilibrium ionization they proposed searching for signatures in line ratios of ion pairs that are populous in the temperature range of interest but have different characteristic lifetimes (e.g. C~IV and O~VI in the transition region, or different ions of Fe at coronal temperature). \cite{Bradshaw2004} investigated non-equilibrium ionization in a small compact flare, using the same numerical model as \cite{Bradshaw2003b}, and localised the energy release in the corona to drive the flare evolution by thermal conduction. During the impulsive phase they found the emissivities of He~I, He~II and C~IV in the transition region to be strongly enhanced above their expected equilibrium values, which was then followed by a significant reduction leading to an increase in the amount of chromospheric plasma ablated into the corona (less energy radiated in the transition region leaves more energy available to drive ablation). During the initial energy release the charge state of the coronal ions was seen to evolve substantially out-of-equilibrium with the increasing temperature and line ratio measurements would yield plasma temperatures that are much greater than the formation temperature of the emitting ion. During the gradual phase the emissivity at transition region temperatures was suppressed releative to equilibrium with reduced downflow velocities, since the enthalpy flux did not have to work as hard to power the transition region, and commensurately increased radiative cooling time-scales. The flare emission as it would be detected by TRACE in its 171~\AA~and 195~\AA~wavelength bands was computed and it was found that the filter ratio technique can give reasonably good estimates of the plasma temperature in quiescence. However, when the populations of Fe~VIII, Fe~IX, Fe~X and Fe~XII exhibited non-equilibrium effects the temperatures derived from filter ratio measurements were unreliable.

\begin{figure}
\centering
\includegraphics[width=1.0\textwidth]{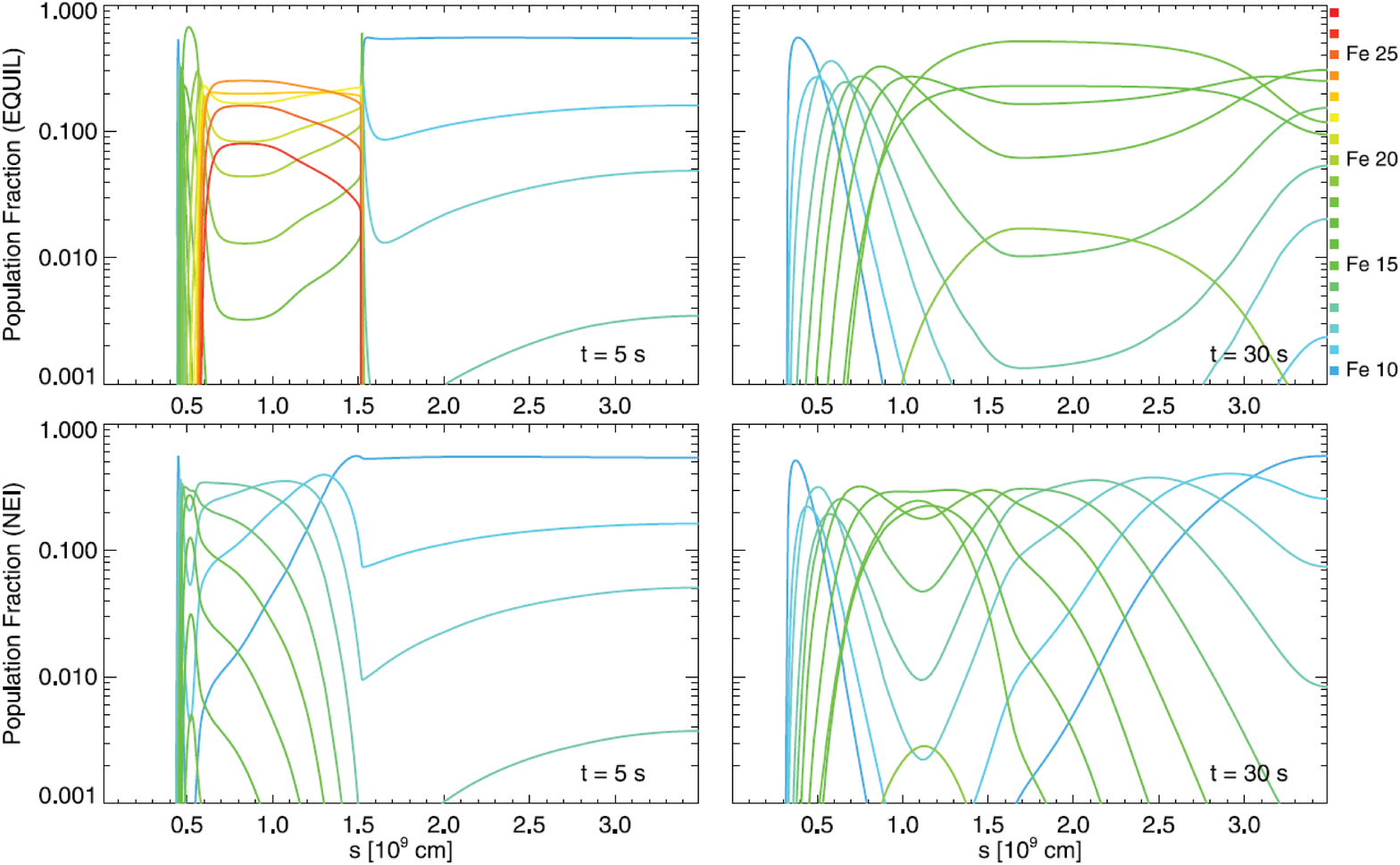}
\caption{Ion population fractions along a heated coronal loop calculated in equilibrium and non-equilibrium. The highly charged states associated with the hottest plasmas in equilibrium are never reached in the non-equilibrium calculation before the onset of cooling. Image credit: \cite{Reale2008}.}
\label{fig12}
\end{figure}

\cite{Bradshaw2006} and \cite{Reale2008} considered strong or `explosive' heating, on short timescales to high temperatures, in an initially rarefied coronal loop atmosphere, to determine the consequences for the evolution of the ionization state. They found extremely strong departures from ionization equilibrium and concluded that for sufficiently short heating events the charge states characteristic of the highest temperatures reached ($10-30$~MK) could never be created before the onset of fast cooling by thermal conduction and coronal filling by chromospheric ablation (Figure~\ref{fig12}). In consequence, the emission measure peaks at temperatures significantly lower than the peak temperature of the plasma and forward modeling emission in the wavelength range of Hinode-EIS showed that no `hot' (e.g. $>10$~MK) component of the plasma would be detected. Heating models that assume ionization equilibrium predict such a hot component, but no observational evidence has yet been found and so non-equilibrium ionization presents one possibility to reconcile observations with current theory. In the future, observations of the solar X-ray continuum could be used to confirm, or otherwise, the presence of a hot component to the emission. The X-ray continuum is due mostly to bremsstrahlung from H and He, which is not sensitive to non-equilibrium ionization effects. There is a significant contribution from radiative recombination in over-ionized plasmas and radiative recombination continua (RRC) are seen in a few X-ray binaries \citep[Cyg X-3:][]{Paerels2000} and supernova remnants \citep[IC~443:][]{Yamaguchi2009}. In under-ionized plasmas, such as may be created in the case of rapid heating in the solar corona, the RRC is weak compared to the bremsstrahlung. In any case the RRC scales as $\exp(h\nu/k_BT)$ and so provided the edges of the most abundant elements are avoided, the continuum shape can be used to diagnose the temperature. \cite{Bradshaw2011} conducted a more extensive survey of the parameter space of energy release magnitudes and timescales, and carried out forward modeling to predict the emission that would be detected in the passbands of the recently launched SDO-AIA, in order to perform a more detailed evaluation of the potential for non-equilibrium ionization to explain the high temperature part of the emission measure. The study led to several conclusions: (1) Deviations from equilibrium were found to be greatest for short-duration nanoflares at low initial coronal densities. (2) Hot emission lines were the most affected and could be suppressed to the point of invisibility. (3) For many of the heating scenarios considered the emission detected in several of the SDO-AIA channels (131, 193, and 211~\AA) was predicted to be dominated by warm, overdense, cooling plasma. (4) It was found to be difficult to avoid creating coronal loops that emit strongly at 1.5 MK and in the range 2-6 MK, which are the most commonly observed kind, for a broad range of nanoflare scenarios; the mere abundance of such loops does not help to constrain the heating parameter space. (5) The Fe XV (284.16~\AA) emission predicted by most of the models was about 10 times brighter than the predicted Ca XVII (192.82~\AA) emission, consistent with observations. \cite{Bradshaw2011} concluded that small-scale, impulsive heating that induces non-equilibrium ionization leads to predictions for observable quantities that are entirely consistent with what is actually observed. 

On larger spatial scales \cite{Rakowski2007} examined the ionization state of several elements derived from in-situ observations of a halo coronal mass ejection (CME) made by the Advanced Composition Explorer (ACE). They assumed an evolution for the CME based on observations and models, and solved the ion population equations for the elements to be compared with the ACE data. They found that plasma in the core of the CME required further heating, possibly due to post-eruptive reconnection following the filament eruption, to reconcile the predicted and observed populations. Plasma in the CME cavity, however, was found not to be further ionized following the eruption, because the low density in that region effectively freezes the ion populations in the state they existed in close to the Sun. \cite{Murphy2011} and \cite{landi10} used the time-dependent ionization in CME ejecta to constrain the temperature history of the expanding plasma and show that an amount of heat comparable to the kinetic energy must be injected to counteract the radiative and adiabatic expansion cooling. \cite{ko10} examined the time-dependent ionization in post-eruption current sheets for a Petschek-type reconnection exhaust, and found that the observable line intensities depended strongly on the height of the reconnection X-line, as well as the density and magnetic field in the surrounding plasma. 

\subsection{Non-Maxwellian Distributions}

The heat flux is sensitive to the underyling distribution of the particles that carry it and since properties of the distribution may also manifest in the emission spectra, then one may consider predicting observable signatures of the heat flux based on the results of numerical experiments and then searching for them in observational datasets. \cite{Karpen1989} used the results of their earlier flare calculations \citep{Karpen1987} to investigate the effect of different heat flux formulations on the X-ray resonance lines of Ca~XIX and Mg~XI and included non-equilibrium ionization in their computation of the spectral lines. By comparing the results of spectroscopic diagnostics carried out with the predicted and observed emission lines, they found the properties of the flare plasma to be most consistent with the non-local formulation of the heat flux. \cite{Jiang2006} found thermal conductivity supressed relative to the classical value in a loop-top source during the late decay phase of a flare, strongly indicative of the onset of flux limiting. The measured cooling timescale was longer than that predicted by classical thermal conduction, but shorter than for radiative cooling. However, they were unable to definitively determine whether plasma wave turbulence was providing additional heating and~/~or surpressing conduction by scattering electrons.

\cite{Esser2000} address the issue that electron temperatures observed at the solar wind acceleration site in the inner corona are too low to give rise to the ion populations observed in-situ in the solar wind, by considering non-Maxwellian electron distributions. They show that reconciling the low electron temperatures and the relatively highly charged ions requires a number of conditions to be satisfied in the inner corona. (1) The electron distribution function must be near-Maxwellian at the coronal base. (2) A departure from Maxwellian must then occur rapidly as a function of height, reaching essentially interplanetary properties within a few solar radii. (3) Ions of different elements must have different speeds to separate their freezing-in distances enough that they encounter different distributions.They also show that the required distributions are very sensitive to the electron temperature, density, and ion flow speed profiles in the coronal region where the ions form. 

A number of studies have adopted forms of non-Maxwellian electron distributions to explain differences between the predicted and observed properties of emission lines. \cite{Dufton1984} found that discrepancies between the observed and theoretically predicted ratios of lines from Si~III could plausibly be explained by non-Maxwellian electron distributions. However, \cite{anderson} showed that the $\kappa$=2.5 distribution used in velocity filtration models of coronal heating overpredicts the intensities of lines normally formed near $10^5$ K by a factor of 100. \cite{Pinfield1999} presented evidence that observations made by SoHO-SUMER showing enhancements in the predicted Si~III 1313~\AA~line intensity by a factor of 5 in active regions, and by a factor of 2 in the quiet Sun and coronal holes, could also be explained by non-Maxwellians. \cite{Ralchenko2007} showed that the excess brightness of some hotter lines (low lying transitions in ions formed at temperatures greater than 2~MK) in the quiet corona may be accounted for by a two-component Maxwellian, where a high-energy component adding 5\% electrons in the temperature range $300-400$~eV is needed to account for the excess brightness.

\cite{Dzifcakova2008c} carried out thermal and non-thermal diagnostics of a solar flare observed with RESIK and RHESSI. They found that in comparison with a synthetic isothermal or multithermal spectra, a non-thermal synthetic spectrum fitted the observed Si~XII dielectronic satellite lines much more closely (with error less than 10\%), and concluded that evidence for significant deviations of the free electron distribution from Maxwellian during the impulsive phase of a solar flare can be diagnosed using X-ray spectral observations. \cite{Dzifcakova2011a} explained features of the RESIK X-ray flare spectra using a Maxwellian or $n$-distribution for the bulk and a power-law tail, finding that the power-law tail has only a small effect on the satellite-to-allowed Si~XIId~/~Si~XIII ratio, which is sensitive to the shape of the bulk distribution and allows the parameter $n$ to be diagnosed. \cite{Kulinova2011} carried out diagnostics of non-thermal distributions in solar flares observed with RESIK and RHESSI. They used two independent diagnostic methods, both indicating the flare plasma affected by the electron beam can have a non-thermal component in the $2–5$~keV range, which was found to be well-described by the $n$-distribution. Spectral line analysis also revealed that the $n$-distribution does not occupy the same spatial location as the thermal component detected by RHESSI at 10~keV. \cite{Karlicky2012} investigated the physical meaning of $n$-distributions in solar flares. The electron component of the return current in a beam-plasma system was shown to have the form of a moving Maxwellian and this was found to be very similar to the high-energy part of an $n$-distribution.

\cite{Dudik2009} calculated TRACE EUV filter responses to emission arising from non-Maxwellian distributions and showed that for $\kappa$-distributions the resulting responses to emission are more broadly dependent on temperature, and their maxima are flatter than for the Maxwellian electron distribution. \cite{Dzifcakova2010} computed a set of synthetic spectra for various $\kappa$-distributions with varying electron densities and mean energies in the spectral range corresponding to the Hinode/EIS and Coronas-F/SPIRIT detectors. Strong EUV lines of Fe in various degrees of ionization were used to analyze the sensitivity of the line ratios to the shape of the distribution function, electron density, and temperature. It was found that EUV coronal Fe lines are generally not very suitable for diagnosing the non-thermal distributions due to their high sensitivity to electron density, but pairs of Fe XVII lines were reasonably good candidates for non-thermal diagnostics. Finally, \cite{Dzifcakova2011b} was able to explain the observed intensity of the Si~III spectrum in coronal holes, the quiet Sun and active region transition regions by adopting an underlying $\kappa$-distribution for the electrons.

\section{Summary and Future Directions}
\label{summary}

We have reviewed a number of the microphysical processes occuring in optically-thin, astrophysical plasma environments, such as supernova remnants and the solar corona, that specifically influence their internal physics such as energy transport and atomic processes and, in consequence, their emission spectra. In particular, we have described the theory of spectral line formation in the coronal approximation and how it is affected by the de-coupling of the ion population from the local temperature (non-equilibrium ionization), that arises when collisional processes are unable to keep pace with heating or cooling, and by the formation of non-Maxwellian particle distribution functions, which are also related to the collisionality of the plasma. Calculations of the charge state of the plasma both in and out of equilibrium, and the most computationally tractable formulations of the kinetic equations that are solved to find the underlying particle distribution function have been presented. A selection of methods have been discussed by which the heat flux can be calculated in a hot plasma, when even near-thermal electrons have long mean-free-paths in relation to the characteristic spatial scales of the system, inducing strongly non-Maxwellian distributions, without recourse to solving a full kinetic equation. In addition, the ways in which non-Maxwellian distributions alter the rates of excitation and ionization have been considered. The results and findings from practical applications of these calculations have been shown throughout and the physics of these processes has been connected with the total radiative energy loss from the system. Finally, evidence for observational signatures of non-equilibrium ionization and non-Maxwellian particle distributions has been presented in association with discussions of the diagnostics that have been used to reveal their influence.

In the future, as astrophysical plasmas are probed with ever greater spatial, temporal and spectral resolution, we expect the microphysical processes that we have discussed here to become increasingly important to developing a full understanding of the physics that drives and governs these systems. The forthcoming Interface Region Imaging Spectrograph \citep[IRIS:][]{DePontieu2009} will provide detailed spectroscopic observations of exactly the region of the Sun's atmosphere where non-equilibrium ionization and non-Maxwellian particle distributions may play a large role in forming the spectral emission lines and, consequently, in determining what information about those regions can be extracted from the spectra by diagnostic studies. The upper-chromosphere and lower-transition region are highly dynamic environments where non-equilibrium ionization has been predicted to be a significant factor in emission from ions such as C~IV and Si~IV (Li-like and Na-like, respectively) that are undergoing heating \citep{Judge2012}; their enhanced emission would lead to over-estimates of the density if equilibrium ionization were assumed when interpreting the observations. IRIS may be able to shed light on the strength of departures from equilibrium in this regime.

Furthermore, streaming particles that enhance the tails of the particle distributions in the interface region may play a role in producing emission from ions of higher charge state than would be predicted from the local temperature alone. A larger tail population provides more electrons with sufficient energy to ionize the ambient plasma to a greater degree. The source of the streaming particles may be a hot ($\ge 10$~MK) component of the coronal emission due to in-situ heating in a high-altitude region where the energy per particle is large, leading to high temperatures and mean-free-paths of lengths on the order of the spatial scales of the magnetic structures, even for near-thermal electrons. Observations of the corona at the highest spatial resolution so far achieved \citep[75~km:][]{Cirtain2013} indicate the presence of entwined bundles of magnetic flux that may be reminiscent of the long-theorized braiding of magnetic field lines leading to reconnection and heating \citep[e.g.][]{Parker1983}. The hot component may be extremely difficult, if not impossible, to observe directly if the plasma is initially tenuous \citep{Bradshaw2006,Reale2008}, but if reconnection does lead to high temperatures and streaming particles, or direct particle acceleration, in the corona, then suitably sophisticated numerical models may be able to predict detectable signatures from lower altitude emission as indirect evidence that can be searched for in real observations by instruments such as IRIS.

Finally, the initial state of what ultimately becomes coronal plasma after heating occurs is another question that is worthy of attention. For example, does heating occur during active region emergence or following the draining of the material dredged up from below the surface as it rises? If the plasma carried to high altitudes cools below $\approx$~20,000~K then it becomes significantly partially ionized and, no longer supported by the Lorentz force due to the emerging field, the neutral atoms rain back onto the surface. In this scenario one might expect the active region plasma to be in an initially tenuous state and the energy per particle relatively high in the case of direct heating, giving rise to very high temperatures where the magnetic field strength and free energy are greatest (in the core of the active region). This may also be the case if the heating is intermittent and the corona is allowed to drain substantially between heating events. Evidence is beginning to accumulate to suggest that the frequency of heating in active regions increases with its age \citep{UgarteUrra2012}; young active regions are heated by low-frequency events \citep[e.g.][]{MuluMoore2011,Bradshaw2012} and older active regions are heated with greater frequency \citep[e.g.][]{Warren2010,Reep2013}. The physics of heat flux saturation and non-local thermal conduction must come into play when considering the energy transport and the overall energy balance of a hot but tenuous atmosphere and the treatments that extend the classical heat flux, described in Section~\ref{heatflux}, must be revisited. In the case of steady heating, where the atmosphere is near hydrostatic, the flux saturation regime is not reached but non-local thermal conduction may still be important in the high-temperature cores of active regions.

\begin{acknowledgement}
The authors would like to thank Dr. Helen Mason for comments on Sections~\ref{otemlines} and \ref{ionization} and Professor Peter Cargill for his comments and suggestions for improving the manuscript as a whole. We also thank the anonymous referee for their valuable input.
\end{acknowledgement}

\bibliographystyle{aps-nameyear}      
\bibliography{example}   
\nocite{*}


\end{document}